\documentclass[12pt,english,sort&compress, letterpaper, ]{JHEP3}
\usepackage[T1]{fontenc}
\usepackage[latin9]{inputenc}
\usepackage{amsthm}
\usepackage{amsmath}
\usepackage{amssymb}
\usepackage[numbers]{natbib}

\makeatletter
\title{Heterotic Flux Attractors}
\author{Lilia Anguelova \\ 
Department of Physics, University of Cincinnati,\\
Cincinnati, OH 45221, USA \\
E-mail: \email{anguella@ucmail.uc.edu}}
\author{Finn Larsen and Ross O'Connell \\ 
Department of Physics, University of Michigan,\\
450 Church Street, Ann Arbor, MI 48109-1020, USA \\ 
E-mail: \email{larsenf@umich.edu}, \email{rcoconne@umich.edu}}
\preprint{MCTP-10-23}
\abstract{
We find attractor equations describing moduli stabilization for heterotic
compactifications with generic $SU(3)$-structure. Complex structure and K\"{a}hler moduli 
are treated on equal footing by using $SU(3)\times SU(3)$-structure at intermediate steps. 
All independent vacuum data, including VEVs of the stabilized moduli, is encoded in a 
pair of generating functions that depend on fluxes alone. We work out an explicit example
that illustrates our methods.  
}

\makeatother

\usepackage{babel}

\makeatother

\usepackage{babel}

\makeatother

\usepackage{babel}

\begin{document}
\global\long\def\hpa{\hphantom{A}}

\section{Introduction}

Geometric moduli are an intriguing aspect of many 4D string vacua.
In simple geometric compactifications, these scalar fields are massless.
The construction of realistic string vacua therefore requires the
addition of new ingredients that stabilize the moduli. In the context
of Type II compactifications much progress can be made by the addition
of both RR and NS fluxes \citep{Dasgupta:1999ss,Giddings:2001yu,DeWolfe:2005uu}.
On the other hand, in heterotic compactifications there are no RR
fluxes and so the problem of moduli stabilization appears more challenging.

In \citep{Becker:2003yv,Becker:2003gq} it was argued that the Kähler
moduli that are left unfixed by the NS flux superpotential \citep{Becker:2002jj},
can be stabilized by considering more general internal manifolds that
still preserve $\mathcal{N}=1$ supersymmetry, namely manifolds with $SU(3)$-structure
instead of $SU(3)$-holonomy. Such non-Kähler compactifications were
first considered in \citep{Strominger:1986uh} and the supersymmetry
conditions, derived in that work, were translated to the modern language
of $SU(3)$-structures in \citep{Cardoso:2002hd}. The deviation from
$SU(3)$-holonomy can be viewed as a kind of geometric flux and thus,
naturally, leads to the generation of a superpotential for the Kähler
moduli \citep{Becker:2003gq,Cardoso:2003af,Gurrieri:2004dt}. One
can study the resulting scalar potential with standard methods, as
in \citep{deCarlos:2005kh} for example. However, here we will use
an alternative method, that provides a powerful systematic tool to
address the problem of moduli stabilization. Instead of directly
minimizing the relevant scalar potential, we will study a set of flux attractor equations.

Attractor equations for flux vacua were first proposed for CY orientifolds
in type IIB \citep{Kallosh:2005ax}, based on similarities with black
hole attractors in supergravity \citep{Ferrara:1995ih,Ferrara:1996dd,Strominger:1996kf}.
Subsequently, this method was extended to encompass much broader classes
of compactifications \citep{DallAgata:2006nr,Bellucci:2007ds,Anguelova:2008fm,Cassani:2009na,Larsen:2009fw,Larsen:2009tv,Anguelova:2009az,Kimura:2008tq}.
The flux attractor equations provide a set of \emph{algebraic} relations
between the fluxes and the stabilized values of the moduli. These
are analogous to the black hole attractor equations, which relate
the black hole charges and the values of the moduli at the horizon.

In order to derive the heterotic attractor equations, we will use
the $SU(3)\times SU(3)$-structure formalism \citep{Grana:2005ny,Grana:2005sn,Grana:2006hr,Grana:2008yw,Cassani:2007pq}.
This formalism is a very efficient tool for studying $\mathcal{N}=2$
compactifications in type IIA/B and their $\mathcal{N}=1$ orientifolds.
In particular, it allows all form-fluxes, geometric fluxes and a set
of non-geometric fluxes to be neatly encoded in a single {}``charge''
matrix $\mathbb{Q}$. 
While the most general $\mathcal{N}=1$ heterotic compactifications are on manifolds with $SU(3)$-structure \citep{Strominger:1986uh}, we will find it rather convenient and illuminating to use the full $SU(3)\times SU(3)$-structure because this
allows symmetric treatment of the complex structure and Kähler moduli, and makes apparent the similarity with the IIB case. A beautiful 
structure will then remain after restriction to the diagonal $SU(3)$
subgroup that describes heterotic $\mathcal{N}=1$ flux attractors.
In particular, such a restriction involves setting to zero the extra components 
of the charge matrix $\mathbb{Q}$. The resulting heterotic attractor equations
extend those of \citep{DallAgata:2006nr,Anguelova:2008fm} 
in a number of interesting ways, as will become clear below.

An intriguing and potentially rather important aspect of the flux
attractor equations is that they can be simplified further by the
introduction of a generating function \citep{Larsen:2009fw,Larsen:2009tv}.
Recall that for black hole attractors the generating function is the
entropy of the black hole \citep{Ooguri:2004zv}. Thus, it is
natural to expect that in the present context the generating
function could also have a deeper meaning. Understanding this could have
profound implications for the string landscape and the kinds of predictions
that it leads to for four-dimensional physics. With this in mind,
we will also 
develop
generating functions for heterotic attractors. 
Recall that in the IIB case derivatives of the generating
function with respect to the fluxes reproduce the stabilized values
of the moduli, as well as their masses \citep{Larsen:2009fw}. In
the heterotic case we will find \emph{two} generating functions, one
of which governs the complex structure moduli and the other governs
the Kähler moduli. A surprising result is that these generating functions
are very closely related to those arising in IIB compactifications.
Indeed, solutions of the IIB attractor equations could be used as starting points for solutions of the heterotic attractor equations.

Finally, we will illustrate the utility of our approach by 
studying an example with one complex structure modulus and one Kähler modulus.
It is important for our methods that we begin with a generic set of
$H$- and geometric fluxes. However, we nevertheless find it sufficient
to consider a simplified flux configuration, and we study moduli stabilization
explicitly in that context. We find the values the moduli are fixed
at, as well as their mass parameters. Also, we compute the gravitino
mass and the values of the two generating functions.

This paper is organized as follows. In section \ref{sec:Fluxes-in-Heterotic}
we provide a brief overview of the formalism of $SU(3)\times SU(3)$-structure
compactifications and the flux superpotential for heterotic compactifications.
In section \ref{sec:Attractor-Equations} we write the complete set
of attractor equations and in section \ref{sec:Generating-Functions}
we develop the associated generating functions. Finally, in section
\ref{sec:STU-Model} we use this machinery to solve an explicit 
two-modulus example.

\section{\label{sec:Fluxes-in-Heterotic}Fluxes, $SU(3)$-Structure, and $SU(3)\times SU(3)$-Structures}

In this section we review background material about $SU(3)$- and
$SU(3)\times SU(3)$-structure compactifications and the 4D effective
theories they give rise to. We will keep the exposition very accessible
and introductory; so readers familiar with this formalism may wish
to skip directly to Section \ref{sec:Attractor-Equations}.

Manifolds with $SU(3)$-structure are a natural generalization of
Calabi-Yau manifolds. In particular, they too are characterized by
the existence of a holomorphic 3-form $\Omega$ and a fundamental
2-form $J$. 
However, unlike Calabi-Yau manifolds, generic $SU(3)$-structure manifolds have
$d\Omega\neq0$ and $dJ\neq0$. A further generalization
is provided by backgrounds with $SU(3)\times SU(3)$ structure, which are characterized
by bispinors $\Phi_{+}$ and $\Phi_{-}$. Locally, the
bispinors can be viewed as sums of even and odd forms, respectively.
More precisely, a generic $\Phi_{+}$ is a sum of $0$-, $2$-,
$4$- and $6$-forms, while a generic $\Phi_{-}$ is a sum of $1$-, $3$- and
$5$-forms. Globally, $\Phi_{+}$ and $\Phi_{-}$ are singlets of
$SU(3)\times SU(3)$. Their moduli spaces are special Kähler manifolds
\citep{Hitchin:2002,Grana:2005ny,Grana:2006hr}, just like the moduli
spaces of $J$ and $\Omega$. A notable feature of the $SU(3)\times SU(3)$-structure
formalism is that the special geometries associated with those two
moduli spaces are treated on equal footing. This is the main reason
it will be beneficial for us to study $SU(3)$-structure compactifications
in the language of $SU(3)\times SU(3)$-structure, as we will make
extensive use of special geometry in deriving the attractor equations.

The superpotential for heterotic compactifications on $SU(3)$-structure
backgrounds is \citep{Becker:2003gq,Cardoso:2003af,Gurrieri:2004dt}:
\begin{equation}
W=\int\left[H_{\mathrm{fl}}+d\left(B+iJ\right)\right]\wedge\Omega\,,\label{eq:W-SU3}\end{equation}
 where the NS 3-form has been decomposed as $H=H_{\mathrm{fl}}+dB$.
This can also be written in the following way: \begin{equation}
W=\int\left\langle \mathcal{D}\Phi_{+},\Phi_{-}\right\rangle ,\label{eq:W-SU3xSU3}\end{equation}
 where the bispinors $\Phi_{+}$ and $\Phi_{-}$ are essentially $e^{-\left(B+iJ\right)}$
and $\Omega$, respectively. In addition to the $H$-flux appearing
explicitly in \eqref{eq:W-SU3}, there are geometric fluxes implicit
in the non-vanishing $dJ$ and $d\Omega.$ In \eqref{eq:W-SU3xSU3}.
The operator $\mathcal{D}$ incorporates both the $H$-flux and the
geometric flux, and in generic $SU(3)\times SU(3)$-structure compactifications
$\mathcal{D}$ includes several non-geometric fluxes as well.

In this section we will review how to write \eqref{eq:W-SU3} and
\eqref{eq:W-SU3xSU3} as explicit functions of the moduli and, along
the way, we will also briefly recall the structure of the moduli spaces.
We also review how to parametrize the geometric and non-geometric
fluxes, and how to reduce to $SU(3)$-structure by setting the non-geometric
fluxes to zero.

\subsection{Basis Forms}

In $SU(3)\times SU(3)$-structure compactifications, the moduli of
the 4Deffective theory are obtained by expanding the bispinors $\Phi_{+}$
and $\Phi_{-}$ on an appropriate basis of forms. 
Unfortunately, it is not known in general how to find the appropriate basis forms,
although there are special cases in which they are known explicitly
\citep{KashaniPoor:2007tr,Caviezel:2008ik}. In order to proceed we assume,
as in \citep{Grana:2005ny,Grana:2006hr}, that the internal space
has a finite basis of forms satisfying certain conditions, such that
the resulting effective four-dimensional description is a consistent
gauged supergravity. Let us denote by $\left\{ \alpha_{I},\beta^{I}\right\} $
the odd basis forms, that are locally sums of 1-, 3-, and 5-forms,
and by $\left\{ \omega_{A},\widetilde{\omega}^{A}\right\} $ the set of
even basis forms, that are locally sums of 0-, 2-, 4-, and 6-forms.
Also, let the range of the indices be $I=0,\dots,h_{-}$ and $A=0,\dots,h_{+}$.
Then clearly, we have $h_{+}+h_{-}$ light moduli of which $h_{-}$ are
associated with the deformations of the odd bispinor $\Phi_{-}$,
and $h_{+}$ with the deformations of the even one $\Phi_{+}$. Recall
that for Calabi-Yau compactifications $h_{-}=h^{\left(2,1\right)}$
is the number of complex structure moduli and $h_{+}=h^{\left(1,1\right)}$
is the number of Kähler moduli.

There is a natural product on sums of forms, called the Mukai pairing.
Written in components, it is\begin{eqnarray}
\left\langle \phi_{+},\psi_{+}\right\rangle  & = & \phi_{0}\wedge\psi_{6}-\phi_{2}\wedge\psi_{4}+\phi_{4}\wedge\psi_{2}-\phi_{6}\wedge\psi_{0}\,,\label{eq:Mukai-even}\\
\left\langle \phi_{-},\psi_{-}\right\rangle  & = & -\phi_{1}\wedge\psi_{5}+\phi_{3}\wedge\psi_{3}-\phi_{5}\wedge\psi_{1}\,,\label{eq:Mukai-odd}\end{eqnarray}
 where $\phi_{n}$ refers to the $n$-form component of $\phi_{\pm}.$
The Mukai pairing always gives a sum of 6-forms, suitable for integration
over the entire compact space. The pattern of signs in \eqref{eq:Mukai-even}
and \eqref{eq:Mukai-odd} guarantees that the Mukai pairing is antisymmetric.
This makes it a natural generalization of the wedge product for 3-forms
on 6D manifolds.

When considering pure 3-forms, we can choose a basis that satisfies
\begin{eqnarray}
\int\alpha_{I}^{\left(3\right)}\wedge\beta_{\left(3\right)}^{J} & = & \delta_{I}^{\hpa J}\,,\\
\int\alpha_{I}^{\left(3\right)}\wedge\alpha_{J}^{\left(3\right)}=\int\beta_{\left(3\right)}^{I}\wedge\beta_{\left(3\right)}^{J} & = & 0\,,\end{eqnarray}
 or in matrix notation \begin{equation}
\int\cdot\wedge\cdot\sim\left(\begin{array}{cc}
0 & -1_{n\times n}\\
1_{n\times n} & 0\end{array}\right)\,,\end{equation}
 where $n$ is the number of $\alpha_{I}.$ This is the quadratic
form associated with $Sp\left(2n\right).$ Indeed, this is the same
$Sp\left(2n\right)$ that is a symmetry of the complex structure moduli
space of Calabi-Yau manifolds. A very similar story holds for general
sums of forms under the Mukai pairing. We can choose a basis for our
odd and even basis forms that satisfies\begin{eqnarray}
\int\left\langle \alpha_{I},\beta^{J}\right\rangle  & = & \delta_{I}^{\hpa J}\,,\\
\int\left\langle \alpha_{I},\alpha_{J}\right\rangle =\int\left\langle \beta^{I},\beta^{J}\right\rangle  & = & 0\,,\\
\int\left\langle \omega_{A},\widetilde{\omega}^{B}\right\rangle  & = & \delta_{A}^{\hpa B}\,,\\
\int\left\langle \omega_{A},\omega_{B}\right\rangle =\int\left\langle \widetilde{\omega}^{A},\widetilde{\omega}^{B}\right\rangle  & = & 0\,.\end{eqnarray}
 The matrix representation of the Mukai pairing for the odd forms
is \begin{equation}
\left\langle \cdot,\cdot\right\rangle \sim\left(\begin{array}{cc}
0 & -1_{\left(h_{-}+1\right)\times\left(h_{-}+1\right)}\\
1_{\left(h_{-}+1\right)\times\left(h_{-}+1\right)} & 0\end{array}\right)\,,\end{equation}
 and the representation on the even forms is analogous, with $h_{-}\to h_{+}.$
The two symplectic products have arisen because the spaces of deformations
of the even and odd bispinors $\Phi_{+}$ and $\Phi_{-}$ are both
special Kähler, as we will discuss in the following section.

If we reduce to the case of $SU(3)$-structure, then our basis elements
all become forms of definite degree. The $\left\{ \alpha_{I},\beta^{I}\right\} $
are all 3-forms, while the $\omega_{A}$ are split into $\omega_{0},$
the unique 0-form, and the 2-forms $\omega_{a},$ $a=1,\dots,h_{+}.$
The $\widetilde{\omega}^{A}$ are split into the 4-forms $\widetilde{\omega}^{a}$
and 6-form $\widetilde{\omega}^{0},$ satisfying \begin{eqnarray}
\kappa_{abc}\widetilde{\omega}^{a} & = & \omega_{b}\wedge\omega_{c}\,,\\
\kappa_{abc}\widetilde{\omega}^{0} & = & \omega_{a}\wedge\omega_{b}\wedge\omega_{c}\,,\end{eqnarray}
 where the $\kappa_{abc}$ are analogue of the CY triple intersection
numbers.

\subsection{Moduli Space}

As already noted, the moduli space of an $SU(3)\times SU(3)$ structure
compactification is the product of two special Kähler manifolds, parameterized
by the deformations of the two bispinors $\Phi_{+}$ and $\Phi_{-}$.
Let us expand the latter on the basis of forms we introduced in the
previous subsection: \begin{eqnarray}
\Phi_{-} & = & Z^{I}\alpha_{I}-F_{I}\beta^{I}\,,\label{eq:Phi-expansion}\\
\Phi_{+} & = & X^{A}\omega_{A}-G_{A}\widetilde{\omega}^{A}\,.\label{eq:Phi+expansion}\end{eqnarray}
 Note that not all of the expansion coefficients are independent.
More precisely, the $F_{I}$ $\left(G_{A}\right)$ are holomorphic
functions of the $Z^{I}$ $\left(X^{A}\right)$ and are homogeneous
of degree one in those variables. Moreover, they are curl-free \begin{eqnarray}
\frac{\partial F_{I}}{\partial Z^{J}} & = & \frac{\partial F_{J}}{\partial Z^{I}}\equiv F_{IJ}\,,\\
\frac{\partial G_{A}}{\partial X^{B}} & = & \frac{\partial G_{B}}{\partial X^{A}}\equiv G_{AB}\,\end{eqnarray}
 so they are generated by prepotentials, at least locally
 . The $Z^{I}$
and $X^{A}$ are projective coordinates for their respective parts
of the moduli space, with the physical moduli given (in patches where
$Z^{0}\neq0,X^{0}\neq0$) by \begin{eqnarray}
z^{i} & = & \frac{Z^{i}}{Z^{0}}\,,\\
x^{a} & = & \frac{X^{a}}{X^{0}}\,,\end{eqnarray}
 with $i=1,\dots,h_{-}$ and $a=1,\dots,h_{+}.$ The Kähler potential
for each part of the moduli space is: \begin{eqnarray}
K_{\pm} & = & -\log i\int\left\langle \Phi_{\pm},\overline{\Phi}_{\pm}\right\rangle ,\label{eq:K}\\
K_{-} & = & -\log i\left(\overline{Z}^{I}F_{I}-Z^{I}\overline{F}_{I}\right),\label{eq:Km-again}\\
K_{+} & = & -\log i\left(\overline{X}^{A}G_{A}-X^{A}\overline{G}_{A}\right).\label{eq:Kp-again}\end{eqnarray}
 All of these properties are reminiscent of the deformations of the
holomorphic 3-form $\Omega$ of a Calabi-Yau manifold and are due
to special Kähler geometry. The even and the odd parts of moduli space
are treated formally identically, while usually Kähler moduli are
treated rather differently from the complex structure moduli.

It will sometimes be convenient to suppress the basis forms entirely
and write bispinors like $\Phi_{\pm}$ in terms of their components
only. Rather than use a single index to identify the components, we
will use a pair of upstairs and downstairs indices, which are natural
because of the symplectic structure identified above. For example,
we may write \eqref{eq:Phi-expansion} and \eqref{eq:Phi+expansion}
as {}``symplectic doublets'': \begin{eqnarray}
\Phi_{-} & = & \left(\begin{array}{c}
Z^{I}\\
F_{I}\end{array}\right),\label{eq:Phi-plus-vector}\\
\Phi_{+} & = & \left(\begin{array}{c}
X^{A}\\
G_{A}\end{array}\right).\label{eq:Phi-minus-vector}\end{eqnarray}
 Note that because the ranges of the $I$ and $A$ indices are different,
the $\Phi_{-}$ and $\Phi_{+}$ {}``vectors'' have a different number
of components.

When we use the symplectic doublet notation, it is natural to write
the Mukai pairing in terms of a matrix product as well. Let us consider
the integrated Mukai pairing of $\Phi_{-}$ with another sum of odd
forms, $F_{-}=m^{I}\alpha_{I}-e_{I}\beta^{I}.$ This can be written
as: \begin{eqnarray}
\int\left\langle F_{-},\Phi_{-}\right\rangle  & = & F_{-}^{T}\mathbb{S}_{-}\Phi_{-}\\
 & = & \left(m^{J}\, e_{J}\right)\left(\begin{array}{cc}
0 & -\delta_{J}^{\hpa I}\\
\delta_{\hpa I}^{J} & 0\end{array}\right)\left(\begin{array}{c}
Z^{I}\\
F_{I}\end{array}\right)\\
 & = & e_{I}Z^{I}-m^{I}F_{I}\,,\end{eqnarray}
 where we have introduced a matrix which implements the symplectic
product for odd forms: \begin{equation}
\mathbb{S}_{-}\equiv\left(\begin{array}{cc}
0 & -\delta_{J}^{\hpa I}\\
\delta_{\hpa I}^{J} & 0\end{array}\right).\label{eq:Sm-again}\end{equation}
 The analogue for the even forms is:\begin{equation}
\mathbb{S}_{+}\equiv\left(\begin{array}{cc}
0 & -\delta_{B}^{\hpa A}\\
\delta_{\hpa A}^{B} & 0\end{array}\right).\end{equation}

Finally, let us comment on how the above considerations specialize
to $SU(3)$-structure. In this case, the bispinors $\Phi_{+}$ and
$\Phi_{-}$ are related to the holomorphic 3-form and complexified
Kähler form via:\begin{eqnarray}
\Phi_{-} & = & \Omega_{3}\,,\label{eq:Phi-SU3}\\
\Phi_{+} & = & X^{0}e^{-\left(B+iJ\right)}\,.\label{eq:Phi+SU3}\end{eqnarray}
 Note that both $\Phi_{+}$ and $\Phi_{-}$ are sections of complex
line bundles. This is why in the $SU(3)$-structure case we have included
explicitly the fiber $X^{0}$ for the Kähler moduli space. Now, substituting
\eqref{eq:Phi+SU3} into \eqref{eq:K} and setting $\left|X^{0}\right|=1$,
we recover the usual Kähler potential \begin{equation}
K_{+}^{SU(3)}=-\log\frac{4}{3}\int J\wedge J\wedge J\,.\end{equation}

\subsection{\label{sub:Fluxes}Fluxes}

On a space with $SU(3)\times SU(3)$-structure, one can introduce
a generalization $\mathcal{D}$ of the exterior derivative $d$, which
still maps odd forms to even ones and vice versa, but does not necessarily
increase the degree of a form by one, even locally. $\mathcal{D}$
is determined by its action on the basis elements $\left\{ \alpha_{I},\beta^{I}\right\} $
and $\left\{ \omega_{A},\widetilde{\omega}^{A}\right\} ,$ 
\begin{equation}
\mathcal{D}\left(\begin{array}{c}
\beta^{I}\\
\alpha_{I}\end{array}\right)=\left(\begin{array}{cc}
q_{A}^{\hpa I} & -q^{AI}\\
-q_{AI} & q_{\hpa I}^{A}\end{array}\right)\left(\begin{array}{c}
\widetilde{\omega}^{A}\\
\omega_{A}\end{array}\right).\label{eq:D-odd}
\end{equation}
 This is essentially a passive picture, since we are acting on the
basis elements. We can also work in an active picture, where $\mathcal{D}$
is represented as a rectangular matrix $\mathbb{Q}$ acting on the
\emph{components} of $\Phi_{\pm}:$ \begin{eqnarray}
\mathcal{D}\Phi_{-}=\mathbb{Q}\Phi_{-}=\left(\begin{array}{cc}
q_{\hpa I}^{A} & q^{AI}\\
q_{AI} & q_{A}^{\hpa I}\end{array}\right)\left(\begin{array}{c}
Z^{I}\label{qcomp}\\
F_{I}\end{array}\right)=\left(\begin{array}{c}
q_{\hpa I}^{A}Z^{I}+q^{AI}F_{I}\\
q_{AI}Z^{I}+q_{A}^{\hpa I}F_{I}\end{array}\right).\label{eq:Q-again}\end{eqnarray}
 The uncontracted indices on the righthand side are $A$'s, not $I$'s,
so $\mathcal{D}\Phi_{-}$ has the same number of components as $\Phi_{+}.$
The active and passive approaches both give\begin{equation}
\mathcal{D}\Phi_{-}=\left(q_{\hpa I}^{A}Z^{I}+q^{AI}F_{I}\right)\omega_{A}-\left(q_{AI}Z^{I}+q_{A}^{\hpa I}F_{I}\right)\widetilde{\omega}^{A}\,.\end{equation}
 We will primarily use the passive picture in order to map the components
of $\mathbb{Q}$ to the usual parametrizations of the fluxes, and
will primarily use the active picture when developing the attractor
equations.
 
The action of  $\mathcal{D}$ on sums of even forms is determined by requiring consistency of integration by parts, i.e. that $\int\left\langle \mathcal{D}A_{-},B_{+}\right\rangle =\int\left\langle A_{-},\mathcal{D}B_{+}\right\rangle $
for all odd forms $A_{-}$ and even forms $B_{+}$. In the passive picture this requires that 
\begin{equation}
\mathcal{D}\left(\begin{array}{c}
\widetilde{\omega}^{A}\\
\omega_{A}\end{array}\right)=\left(\begin{array}{cc}
q_{\hpa I}^{A} & q^{AI}\\
q_{AI} & q_{A}^{\hpa I}\end{array}\right)\left(\begin{array}{c}
\beta^{I}\\
\alpha_{I}\end{array}\right).
\end{equation}
In the active picture we find that 
\begin{eqnarray}
\int\left\langle \mathcal{D}A_{-},B_{+}\right\rangle  & = & A_{-}^{T}\mathbb{Q}^{T}\mathbb{S}_{+}B_{+}\\
 & = & A_{-}^{T}\mathbb{S}_{-}\mathbb{S}_{-}^{T}\mathbb{Q}^{T}\mathbb{S}_{+}B_{+}\\
 & = & \int\left\langle A_{-},\mathbb{S}_{-}^{T}\mathbb{Q}^{T}\mathbb{S}_{+}B_{+}\right\rangle ,
 \end{eqnarray}
Again requiring consistency of integration by parts, we find that 
 \begin{eqnarray}
\mathcal{D}B_{+} & = & \mathbb{S}_{-}^{T}\mathbb{Q}^{T}\mathbb{S}_{+}B_{+}\,,\label{eq:Deven}\\
\mathbb{S}_{-}^{T}\mathbb{Q}^{T}\mathbb{S}_{+} & = & \left(\begin{array}{cc}
q_{A}^{\hpa I} & -q^{AI}\\
-q_{AI} & q_{\hpa I}^{A}\end{array}\right)
\end{eqnarray}
 for any sum of even forms $B_{+}.$


In a generic $SU\!\left(3\right)\times SU\!\left(3\right)$-structure
background, $\mathcal{D}$ contains several operators. One is the
exterior derivative $d$, which increases the degree of a form by
one in a way parametrized by the geometric fluxes. Another is $H_{3}\wedge$\,,
which clearly increases the degree of a form by three. Yet another
part of $\mathcal{D}$, often denoted $Q\cdot$, \emph{reduces} the
degree by one and is usually referred to as a non-geometric flux.
Finally, there is also an operator $R\urcorner$, that reduces the
degree of a form by three, and whose status is somewhat more speculative
\citep{Shelton:2005cf,Shelton:2006fd,Grana:2008yw}. In the $SU(3)$
structure case, $\mathcal{D}$ can only \emph{increase} the degree
of a form it acts on \citep{Chiossi:2002,Grana:2005ny}. Hence, specializing
to $SU(3)$-structure is achieved by turning off $Q\cdot$ and $R\urcorner$,
i.e. by allowing only $H$-flux and geometric flux.

The specialization to $SU(3)$-structure is readily implemented on
$\mathbb{Q},$ the matrix representation of $\mathcal{D}$. In the
$SU(3)$-structure case all of our odd basis forms $\left\{ \alpha_{I},\beta^{I}\right\} $
are 3-forms, while half of our even basis forms, namely $\omega_{A}$,
have degree 0 or 2, and the other half, $\widetilde{\omega}^{A}$, have
degree 4 or 6. Thus $\mathcal{D}$ can map $\left\{ \alpha_{I},\beta^{I}\right\} $
to $\left\{ \widetilde{\omega}^{A}\right\} $, but not to $\left\{ \omega_{A}\right\} $.
We implement this by setting $q^{AI}=q_{\hpa I}^{A}=0.$ Also, we
can identify $q_{AI}$ and $q_{A}^{\hpa I}$ with the usual components
of $H_{\mathrm{fl}}$ and geometric fluxes. Let us begin by considering
\begin{eqnarray}
\mathcal{D}\omega_{0} & = & \left(d-H_{\mathrm{fl}}\wedge\right)1=-H_{\mathrm{fl}}=-\left(m_{h}^{I}\alpha_{I}-e_{I}^{h}\beta^{I}\right)\,,\end{eqnarray}
 where $d1=0$, because $SU(3)$-structure compactifications have
no 1-form basis elements. Referring to \eqref{eq:Deven}, we see that
\begin{equation}
\mathcal{D}\omega_{0}=q_{0}^{\hpa I}\alpha_{I}+q_{0I}\beta^{I}\,,\end{equation}
 so we have \begin{eqnarray}
q_{0}^{\hpa I}=-m_{h}^{I}\,\,,\qquad q_{0I}=e_{I}^{h}\,.\label{eq:H-down}\end{eqnarray}
 On the other hand, $q_{a}^{\hpa I}$ and $q_{aI}$ can be identified
with a frequently used parametrization of geometric fluxes \citep{Ihl:2007ah,Larsen:2009tv},
denoted by $r_{a}^{\hpa I}$ and $r_{aI}$ respectively, in the following
way. Consider: \begin{equation}
\mathcal{D}\omega_{a}=\left(d-H_{\mathrm{fl}}\wedge\right)\omega_{a}=d\omega_{a}=-\left(r_{a}^{\hpa I}\alpha_{I}-r_{aI}\beta^{I}\right),\end{equation}
 where $H_{\mathrm{fl}}\wedge\omega_{a}$ vanishes because there are
no 5-form basis elements on $SU(3)$-structure manifolds. Referring
again to \eqref{eq:Deven}, we have\begin{equation}
\mathcal{D}\omega_{a}=q_{a}^{\hpa I}\alpha_{I}+q_{aI}\beta^{I}\,,\end{equation}
 so we identify \begin{eqnarray}
q_{a}^{\hpa I}=-r_{a}^{\hpa I}\,\,,\qquad q_{aI}=r_{aI}\,.\label{eq:geo-down}\end{eqnarray}

Since $\mathcal{D}$ is a generalization of the exterior derivative
$d,$ it is natural to require nilpotency $\mathcal{D}^{2}=0$. In
the general $SU(3)\times SU(3)$ setting this gives rise to the following 
tadpole constraints: \begin{equation}
\left(\begin{array}{cc}
q_{A}^{\hpa I} & -q^{AI}\\
-q_{AI} & q_{\hpa I}^{A}\end{array}\right)\left(\begin{array}{cc}
q_{\hpa J}^{A} & q^{AJ}\\
q_{AJ} & q_{A}^{\hpa J}\end{array}\right)=\left(\begin{array}{cc}
q_{\hpa I}^{A} & q^{AI}\\
q_{AI} & q_{A}^{\hpa I}\end{array}\right)\left(\begin{array}{cc}
q_{B}^{\hpa I} & -q^{BI}\\
-q_{BI} & q_{\hpa I}^{B}\end{array}\right)=0\,.\end{equation}
 These can be rewritten in a variety of ways. In the active picture,
we have \begin{equation}
\mathbb{Q}^{T}\mathbb{S}_{+}\mathbb{Q}=\mathbb{Q}\mathbb{S}_{-}\mathbb{Q}^{T}=0\,.\label{eq:Q-tadpole}\end{equation}
 Expanding out in components, we have the following six conditions:\begin{eqnarray}
q_{A}^{\hpa I}q_{\hpa J}^{A}-q^{AI}q_{AJ} & = & 0\,,\\
q_{A}^{\hpa I}q^{AJ}-q^{AI}q_{A}^{\hpa J} & = & 0\,,\\
q_{AI}q_{\hpa J}^{A}-q_{\hpa I}^{A}q_{AJ} & = & 0\,,\\
q_{\hpa I}^{A}q_{B}^{\hpa I}-q^{AI}q_{BI} & = & 0\,,\\
q_{A}^{\hpa I}q_{BI}-q_{B}^{\hpa I}q_{AI} & = & 0\,,\\
q^{AI}q_{\hpa I}^{B}-q_{\hpa I}^{A}q^{BI} & = & 0\,.\end{eqnarray}
 In the case of $SU(3)$-structure, these are automatically satisfied
except for \begin{equation}
q_{A}^{\hpa I}q_{BI}-q_{B}^{\hpa I}q_{AI}=0\,.\label{eq:SU3-tadpole}\end{equation}

We should note that the above tadpole constraints are valid for the
standard embedding of the 10D gauge field. In more general cases, the heterotic Bianchi identity
\begin{equation}
dH_{\mathrm{fl}}=\frac{\alpha^{\prime}}{4}\left[\mbox{tr}R\wedge R-\mbox{tr}F\wedge F\right]\end{equation}
 would imply additional restrictions since $\mathcal{D}^{2}=-dH_{\mathrm{fl}}$.
A further source for $dH_{\mathrm{fl}}$ can be the presence of NS5-branes.
While considering non-standard embeddings and/or 5-branes is certainly
very interesting and worthy of thorough investigation, we will not pursue the matter here. We will consider only the standard embedding, and apply the standard tadpole constraints
reviewed above.

\subsection{Superpotential and Potential}

In the beginning of this section we stated that the standard superpotential
for heterotic compactifications on $SU(3)$-structure manifolds, namely
\begin{equation}
W=\int\left[H_{\mathrm{fl}}+d\left(B+iJ\right)\right]\wedge\Omega_{3}\,,\label{eq:W-SU3-2}\end{equation}
 can be written as: \begin{equation}
W=\int\left\langle \mathcal{D}\Phi_{+},\Phi_{-}\right\rangle \,,\label{eq:W-SU3xSU3-2}\end{equation}
 in the language of $SU(3)\times SU(3)$-structures. Having in mind
(\ref{eq:Phi-SU3})-(\ref{eq:Phi+SU3}), it is easy to see that this
is the case. Indeed, let us compute $\mathcal{D}\Phi_{+}$ for the
case of $SU(3)$-structure: \begin{eqnarray}
\mathcal{D}\Phi_{+} & = & \left(d-H_{\mathrm{fl}}\wedge\right)X^{0}e^{-\left(B+iJ\right)}\\
 & = & -X^{0}\left[d\left(B+iJ\right)+H_{\mathrm{fl}}\right]\,,\end{eqnarray}
 where again the absence of 1- and 5-form basis elements accounts
for the simplification. It is now obvious that \eqref{eq:W-SU3xSU3-2}
is equivalent to \eqref{eq:W-SU3-2} with the gauge choice $X^{0}=-1.$

One important aspect of the above superpotential is that, while it
depends on all geometric moduli, it does not depend on the dilaton
$S$. Recall that the complexified variable $S$ is given by $S=a+ie^{-2\varphi}$,
where $\varphi$ is the real 4D dilaton and $a$ is the axion that
is 4DHodge-dual to the NS 2-form $B_{\mu\nu}$ with $\mu,\nu$ running
over the external four dimensions. At the classical level the Kähler
potential for $S$ is: \begin{equation}
K_{(S)}=-\log\left(-i(S-\bar{S})\right)\,.\end{equation}
 The scalar potential \begin{equation}
V=e^{K}\left[g^{\alpha\overline{\beta}}D_{\alpha}W\overline{D_{\beta}W}-3\left|W\right|^{2}\right],\end{equation}
 where $\alpha$ runs over $S$ and the Kähler and complex structure
moduli, simplifies to: \begin{equation}
V=e^{K_{+}+K_{-}+K_{(S)}}\left[g^{i\overline{j}}D_{i}W\overline{D_{j}W}+g^{a\overline{b}}D_{a}W\overline{D_{b}W}-2\left|W\right|^{2}\right],\label{eq:V-dilaton}\end{equation}
 since the superpotential does not depend on the dilaton $\partial_{S}W=0$.
On the other hand, $W$ does depend on the Kähler moduli, due to the geometric fluxes 
of the $SU(3)$ structure. So the scalar potential is not positive-definite, 
unlike in the $SU(3)$-holonomy case. This means, in particular, that $V$ can have 
nonsupersymmetric AdS minima.

The above potential can be extremized with respect to the Kähler and
complex structure moduli by imposing the supersymmetry conditions
$D_{i}W=D_{a}W=0$.%
\footnote{The remaining supersymmetry condition, $D_{S}W=0$, clearly implies
$W=0$. We will nevertheless keep the value of $W$ arbitrary, thus
encompassing also AdS vacua with spontaneously broken supersymmetry.%
} We will use the latter to derive our attractor equations. However,
the dilaton remains either a flat direction (when $W=0$) or a run-away direction (when $W\neq 0$). A standard way to stabilize the dilaton is
to take into account gaugino condensation, which leads to an additional
contribution to $W$ of the form $e^{ikS}$ for some constant $k$;
see for example \citep{deCarlos:2005kh}. In the following, however,
we will not consider the dilaton in any detail and will simply assume
that it has been fixed at some value. Developing attractor equations
at the quantum level, i.e. when perturbative and/or non-perturbative
effects are included in the scalar potential, is undoubtedly of great
interest, 
but we leave it for future investigation.

\section{\label{sec:Attractor-Equations}Attractor Equations}

We now develop a set of attractor equations equivalent to the 
F-flatness conditions $D_{i}W=D_{a}W=0.$ We do this by expanding
the flux matrix $\mathbb{Q}$ on a convenient moduli-dependent basis.
The resulting real algebraic equations will simultaneously determine
both the stabilized values of the moduli, and the independent masses
and Yukawa couplings at the attractor point. We also identify a subset
of these equations that are formally identical to the attractor equations
developed for type IIB O3/O7 compactifications in \citep{Larsen:2009fw,Larsen:2009tv}.
These projected equations are naturally written in terms of a set
of complex, moduli-dependent fluxes $\mathcal{D}\Phi_{+}$ and $\mathcal{D}\Phi_{-}.$
They determine all of the moduli, but not all of the masses and Yukawa
couplings.

In our formalism it is quite natural to work with a complete $\mathbb{Q}$
matrix, i.e. one appropriate for generic $SU(3)\times SU(3)$-structure compactifications.
However, we are interested in $\mathcal{N}=1$ compactifications of
the heterotic string, which require that we specialize to $SU(3)$-structure
compactifications\footnote{Describing heterotic $\mathcal{N}=1$ compactifications in the language of $SU(3)\times SU(3)$ structure has been addressed in \cite{Andriot:2009fp} from the 10D perspective. It would be interesting to understand explicitly how to relate their $\mathcal{N}=1$ conditions (4.15) with the 4D effective potential F-flatness conditions that we study here.}. In Section \ref{sub:SU3-AttractorEqs} we will
reduce the $\mathbb{Q}$ matrix to the form appropriate for $SU(3)$-structure
and find the resulting simplification of both the full attractor equations,
and the projected subset.

\subsection{Expanding $\mathbb{Q}$/Change of Variables}

The basic strategy for developing flux attractor equations is to expand
the fluxes in a basis, usually moduli-dependent, where the minimization
conditions are easily implemented. For example, in the case of the
IIB O3/O7 attractor equations the fluxes are encoded in complex 3-forms
and one expands them on the basis $\left\{ \Omega,D_{i}\Omega,\overline{D_{i}\Omega},\overline{\Omega}\right\} $, where $D_i$ is the usual K\"{a}hler derivative.
The minimization conditions then require that a complex combination
of the fluxes be imaginary self-dual, i.e. that the coefficients of
the $\Omega$ and $\overline{D_{i}\Omega}$ terms vanish.

In the present case, we have analogous bases for even and odd sums
of forms, $\left\{ \Phi_{+},D_{a}\Phi_{+},\overline{D_{a}\Phi}_{+},\overline{\Phi}_{+}\right\} $
and $\left\{ \Phi_{-},D_{i}\Phi_{-},\overline{D_{i}\Phi}_{-}\overline{\Phi}_{-}\right\} ,$
but the matrix $\mathbb{Q}$ is neither a sum of even forms nor a
sum of odd forms. Instead, it is a linear map from odd forms to even
forms, i.e. a rectangular matrix, and is properly expanded on $\left\{ \Phi_{+},D_{a}\Phi_{+},\overline{D_{a}\Phi}_{+},\overline{\Phi}_{+}\right\} \otimes\left\{ \Phi_{-},D_{i}\Phi_{-},\overline{D_{i}\Phi}_{-}\overline{\Phi}_{-}\right\} ^{T}.$
More specifically, we consider the following expansion: \begin{eqnarray}
\!\!\!\!\!\!\mathbb{Q} & = & \mbox{Re}\left\{ C\Phi_{+}\Phi_{-}^{T}+C^{i}\Phi_{+}D_{i}\Phi_{-}^{T}+C^{a}\left(D_{a}\Phi_{+}\right)\Phi_{-}^{T}+C^{ai}D_{a}\Phi_{+}\left(D_{i}\Phi_{-}\right)^{T}\right\} {\mathbb{S}}_{-}\nonumber \\
 &  & +\mbox{Re }\!\!\left\{ \widetilde{C}\Phi_{+}\overline{\Phi}_{-}^{T}+\overline{\widetilde{C}}^{\overline{i}}\Phi_{+}\overline{D_{i}\Phi}_{-}^{T}+\widetilde{C}^{a}\left(D_{a}\Phi_{+}\right)\overline{\Phi}_{-}^{T}+\widetilde{C}^{a\overline{i}}D_{a}\Phi_{+}\left(\overline{D_{i}\Phi}_{-}\right)^{T}\right\} {\mathbb{S}}_{-}\,.\label{eq:GeneralQ}\end{eqnarray}
 This expansion can be thought of as a change of basis. Previously,
we specified our choice of fluxes and location in moduli space via
$\left\{ q^{AI},q_{\hpa I}^{A},q_{A}^{\hpa I},q_{AI},z^{i},x^{a}\right\} .$
Equivalently, we can parametrize this data with $\left\{ C,C^{i},C^{a},C^{ai},\widetilde{C},\widetilde{C}^{a},\widetilde{C}^{i}\widetilde{C}^{ai},z^{i}x^{a}\right\} .$
Note that the number of real parameters in $\left\{ C,C^{i},C^{a},C^{ai},\widetilde{C},\widetilde{C}^{a},\widetilde{C}^{i}\widetilde{C}^{ai}\right\} $
is equal to the number of real parameters in $\left\{ q^{AI},q_{\hpa I}^{A},q_{A}^{\hpa I},q_{AI}\right\} .$  This is additional evidence that we have expanded on a complete basis, while previous efforts have not included the second line of \eqref{eq:GeneralQ}.

The trailing factor of $\mathbb{S}_{-}$ in (\ref{eq:GeneralQ}) is
not mandatory for the validity of the expansion but it simplifies
the calculation of Mukai pairings of the form $\left\langle \mathcal{D}U_{-},U_{+}\right\rangle ,$
where $U_{-}$ and $U_{+}$ are sums of forms of odd and even degree,
respectively. Each term in the expansion of $\mathbb{Q}$ is of the
form \begin{equation}
\mathbb{Q}\sim V_{+}V_{-}^{T}\mathbb{S}_{-}\,,\end{equation}
 and for simplicity we will focus on a single term. The Mukai pairings
then factorize as \begin{eqnarray}
\int\left\langle \mathcal{D}U_{-},U_{+}\right\rangle  & = & U_{-}^{T}\mathbb{Q}^{T}\mathbb{S}_{+}U_{+}\\
 & \sim & U_{-}^{T}\mathbb{S}_{-}^{T}V_{-}V_{+}^{T}\mathbb{S}_{+}U_{+}\\
 & = & \int\left\langle U_{-},V_{-}\right\rangle \int\left\langle U_{+},V_{+}\right\rangle .\end{eqnarray}
 We can then exploit the orthogonality properties of the bases $\left\{ \Phi_{+},D_{a}\Phi_{+},\overline{D_{a}\Phi}_{+},\overline{\Phi}_{+}\right\} $
and $\left\{ \Phi_{-},D_{i}\Phi_{-},\overline{D_{i}\Phi}_{-}\overline{\Phi}_{-}\right\} ,$
where the only non-vanishing Mukai pairings are $\!\int\!\left\langle \Phi_{\pm},\overline{\Phi}_{\pm}\right\rangle $
and \begin{eqnarray}
\int\left\langle D_{i}\Phi_{-},\overline{D_{j}\Phi}_{-}\right\rangle  & = & -g_{i\overline{j}}\int\left\langle \Phi_{-},\overline{\Phi}_{-}\right\rangle \,,\\
\int\left\langle D_{a}\Phi_{+},\overline{D_{b}\Phi}_{+}\right\rangle  & = & -g_{a\overline{b}}\int\left\langle \Phi_{+},\overline{\Phi}_{+}\right\rangle \,\end{eqnarray}
 with $g_{i\overline{j}}$ and $g_{a\overline{b}}$ being the metrics
on the odd and even moduli spaces. In all cases that we will consider,
the $U_{-}$ and $U_{+}$ of interest will pick out a single term
from $\mathbb{Q}.$

In order to demonstrate how this works, we compute the superpotential
\eqref{eq:W-SU3xSU3-2}: \begin{equation}
W=-\int\left\langle \mathcal{D}\Phi_{-},\Phi_{+}\right\rangle .\label{eq:W-holo}\end{equation}
 Here $\Phi_{-}$ and $\Phi_{+}$ have taken the role of $U_{-}$
and $U_{+}$ in the previous paragraph and the orthogonality relations
tell us that only the $\overline{\Phi}_{+}\overline{\Phi}_{-}^{T}$
term in \eqref{eq:GeneralQ} will contribute to the righthand side. We therefore
have\begin{eqnarray}
W & = & -\frac{\overline{C}}{2}\int\left\langle \Phi_{-},\overline{\Phi}_{-}\right\rangle \int\left\langle \Phi_{+},\overline{\Phi}_{+}\right\rangle .\label{eq:WC-Real}\end{eqnarray}
 Thus $\overline{C}$ is related to the on-shell value of the superpotential.
Note that the non-holomorphic nature of equation \eqref{eq:WC-Real}
is due to the fact that our change of variables $\left\{ q^{AI},q_{\hpa I}^{A},q_{A}^{\hpa I},q_{AI},z^{i},x^{a}\right\} \to\left\{ C,C^{i},C^{a},C^{ai},\widetilde{C},\widetilde{C}^{a},\widetilde{C}^{i}\widetilde{C}^{ai},z^{i}x^{a}\right\} $
is not holomorphic. In effect, we have traded holomorphy, which is
manifest in \eqref{eq:W-holo}, for good $SU(3)\times SU(3)$ representations;
indeed, $C$ is an $SU(3)\times SU(3)$ singlet, while the $q$'s
of (\ref{qcomp}) are not in good $SU(3)\times SU(3)$ representations.

\subsection{Real Attractor Equations}

While the expansion \eqref{eq:GeneralQ} holds at arbitrary points
in moduli space, it is quite useful when imposing the F-flatness
conditions \begin{eqnarray}
D_{a}W & = & \int\left\langle \mathcal{D}\Phi_{-},D_{a}\Phi_{+}\right\rangle =0\,,\\
D_{i}W & = & \int\left\langle \mathcal{D}D_{i}\Phi_{-},\Phi_{+}\right\rangle =0\,.\end{eqnarray}
 Note that $\mathcal{D}$ acts on the basis forms, while $D_{i}$
and $D_{a}$ act on the expansion coefficients of $\Phi_{-}$ and
$\Phi_{+},$ and so they commute. The $D_{a}W=0$ condition picks
out the $\overline{D_{b}\Phi}_{+}\overline{\Phi}_{-}^{T}$ term in
the expansion of $\mathbb{Q}$, yielding: \begin{equation}
D_{a}W=\frac{\overline{C}^{\overline{b}}}{2}\int\left\langle D_{a}\Phi_{+},\overline{D_{b}\Phi}_{+}\right\rangle \int\left\langle \Phi_{-},\overline{\Phi}_{-}\right\rangle .\end{equation}
 Since the Mukai pairings are non-vanishing, F-flatness requires $C^{b}=0.$
Similar reasoning, applied to $D_{i}W$, implies that $C^{i}=0$ as
well. We therefore arrive at the attractor equation for $\mathbb{Q}:$\begin{eqnarray}
\hspace*{-1em}\mathbb{Q} & = & \mbox{Re}\left\{ C\Phi_{+}\Phi_{-}^{T}+C^{ai}D_{a}\Phi_{+}\left(D_{i}\Phi_{-}\right)^{T}\right\} \mathbb{S}_{-}\nonumber \\
 &  & \!\!\!+\mbox{Re }\!\!\!\left\{ \widetilde{C}\Phi_{+}\overline{\Phi}_{-}^{T}+\overline{\widetilde{C}}^{\overline{i}}\Phi_{+}\overline{D_{i}\Phi}_{-}^{T}+\widetilde{C}^{a}\left(D_{a}\Phi_{+}\right)\overline{\Phi}_{-}^{T}+\widetilde{C}^{a\overline{i}}D_{a}\Phi_{+}\left(\overline{D_{i}\Phi}_{-}\right)^{T}\right\} \mathbb{S}_{-}\,.\label{eq:C-att}\end{eqnarray}
 The appearance of six different terms here may seem discouraging.
Fortunately, this expression can be greatly simplified, as we will
now demonstrate.

We begin by simplifying the second line. Our strategy will be to replace
the Kähler derivatives with ordinary derivatives, $D_{i}\to\partial_{I}$
and $D_{a}\to\partial_{A}.$ The important difference between the
Kähler derivatives and ordinary derivatives is that ordinary derivatives
of $\Phi_{\pm}$ generate terms proportional to $\Phi_{\pm},$ \begin{eqnarray}
\partial_{I}\Phi_{-} & = & K_{I}\Phi_{-}+\dots\,,\label{eq:dI-minus}\\
\partial_{A}\Phi_{+} & = & K_{A}\Phi_{+}+\dots\,,\label{eq:dA-plus}\end{eqnarray}
 while Kähler derivatives do not. This means that $\partial_{A}\Phi_{+}\left(\overline{\partial_{I}\Phi}_{-}\right)^{T}$
generates terms proportional to $\Phi_{+}\overline{\Phi}_{-}^{T},$
$\Phi_{+}\overline{D_{i}\Phi}_{-}^{T},$ and $\left(D_{a}\Phi_{+}\right)\overline{\Phi}_{-}^{T},$
in addition to $D_{a}\Phi_{+}\left(\overline{D_{i}\Phi}_{-}\right)^{T}.$
We can therefore introduce the alternative parametrization \begin{equation}
\widetilde{L}^{A\overline{I}}\partial_{A}\Phi_{+}\left(\overline{\partial_{I}\Phi}_{-}\right)^{T}=\widetilde{C}\Phi_{+}\overline{\Phi}_{-}^{T}+\overline{\widetilde{C}}^{\overline{i}}\Phi_{+}\overline{D_{i}\Phi}_{-}^{T}+\widetilde{C}^{a}\left(D_{a}\Phi_{+}\right)\overline{\Phi}_{-}^{T}+\widetilde{C}^{a\overline{i}}D_{a}\Phi_{+}\left(\overline{D_{i}\Phi}_{-}\right)^{T}.\end{equation}
 Note that $\widetilde{L}^{A\overline{I}}$ constitutes the same number
of complex parameters as $\left\{ \widetilde{C},\widetilde{C}^{i},\widetilde{C}^{a},\widetilde{C}^{a\overline{i}}\right\} .$
After introducing the $\widetilde{L}^{A\overline{I}},$ the attractor
equation (\ref{eq:C-att}) is recast as \begin{equation}
\mathbb{Q}=\mbox{Re}\left\{ C\Phi_{+}\Phi_{-}^{T}+C^{ai}D_{a}\Phi_{+}\left(D_{i}\Phi_{-}\right)^{T}+\widetilde{L}^{A\overline{I}}\partial_{A}\Phi_{+}\left(\overline{\partial_{I}\Phi}_{-}\right)^{T}\right\} \mathbb{S}_{-}\,.\end{equation}

We will apply a variation of this logic to the $C^{ai}D_{a}\Phi_{+}\left(D_{i}\Phi_{-}\right)^{T}$
term. Although this will not reduce the number of terms in the attractor
equation, ordinary derivatives will be significantly easier to work
with than Kähler derivatives. We want to ensure that we do not
generate terms proportional to $\Phi_{+}D_{i}\Phi_{-}^{T}$ and $\left(D_{a}\Phi_{+}\right)\Phi_{-}^{T}$.
This means that, in addition to replacing $C^{ai}D_{a}\Phi_{+}\left(D_{i}\Phi_{-}\right)^{T}\to L^{AI}\partial_{A}\Phi_{+}\left(\partial_{I}\Phi_{-}\right)^{T},$
we must impose two sets of constraints:\begin{equation}
L^{AI}K_{I}=L^{AI}K_{A}=0\,.\label{eq:LK-Constraints}\end{equation}
 While $C^{ai}$ constitutes $h_{+}h_{-}$ complex parameters, $L^{AI}$
constitutes $h_{+}h_{-}+h_{-}+h_{+}+1$ complex parameters. So we
should have $h_{-}+h_{+}+1$ constraints. While it may appear 
that \eqref{eq:LK-Constraints} constitutes $h_{-}+h_{+}+2$ constraints,
the $L^{AI}K_{I}=0$ and $L^{AI}K_{A}=0$ conditions \emph{both} impose
$L^{AI}K_{A}K_{I}=0,$ so they actually constitute only the expected
$h_{-}+h_{+}+1$ constraints. We then have our final form for the
attractor equation, plus constraints:\begin{eqnarray}
\mathbb{Q} & = & \mbox{Re}\left\{ C\Phi_{+}\Phi_{-}^{T}+L^{AI}\partial_{A}\Phi_{+}\left(\partial_{I}\Phi_{-}\right)^{T}+\widetilde{L}^{A\overline{I}}\partial_{A}\Phi_{+}\left(\overline{\partial_{I}\Phi}_{-}\right)^{T}\right\} \mathbb{S}_{-}\,,\label{eq:Q-att}\\
0 & = & L^{AI}K_{I}\,,\label{eq:LK-1}\\
0 & = & L^{AI}K_{A}\,.\label{eq:LK-2}\end{eqnarray}
We can use \eqref{eq:Phi-plus-vector}, \eqref{eq:Phi-minus-vector}, \eqref{eq:Sm-again}, and \eqref{eq:Q-again} to expand \eqref{eq:Q-att} in terms of components, and use \eqref{eq:Km-again} and \eqref{eq:Kp-again} to evaluate $K_I$ and $K_A$:
 \begin{eqnarray}
-q^{AI} & = & \mbox{Re}\left\{ CX^{A}Z^{I}+L^{AI}+\widetilde{L}^{A\overline{I}}\right\} ,\label{eq:RealAtt-1}\\
q_{\hpa I}^{A} & = & \mbox{Re}\left\{ CX^{A}F_{I}+L^{AJ}F_{IJ}+\widetilde{L}^{A\overline{J}}\overline{F}_{\overline{IJ}}\right\} ,\label{eq:RealAtt-2}\\
-q_{A}^{\hpa I} & = & \mbox{Re}\left\{ CG_{A}Z^{I}+L^{BI}G_{AB}+\widetilde{L}^{B\overline{I}}G_{AB}\right\} ,\label{eq:RealAtt-3}\\
q_{AI} & = & \mbox{Re}\left\{ CG_{A}F_{I}+L^{BJ}G_{AB}F_{IJ}+\widetilde{L}^{B\overline{J}}G_{AB}\overline{F}_{\overline{IJ}}\right\} ,\label{eq:RealAtt-4}\\
0 & = & L^{AI}\left(\overline{F}_{I}-\overline{Z}^{J}F_{IJ}\right),\label{eq:LConstraint-1}\\
0 & = & L^{AI}\left(\overline{G}_{A}-\overline{X}^{A}G_{AB}\right).\label{eq:LConstraint-2}\end{eqnarray}
 Note that the number of real equations, $4\left(h_{-}+1\right)\left(h_{+}+1\right)+2\left(h_{-}+h_{+}+1\right)$,
is equal to the number of variables in $\left\{ z^{i},x^{a},CX^{0}Z^{0},L^{AI},\widetilde{L}^{A\overline{I}}\right\} .$

\subsection{\label{sub:Complex-Attractor-Equations}Projected Attractor Equations}

The benefit of (\ref{eq:RealAtt-1}-\ref{eq:LConstraint-2}) is that
every flux and every independent parameter of the compactification
appears explicitly. We now identify a \emph{subset} of these equations
that is easier to solve. Instead of working in terms of the real flux
matrix $\mathbb{Q},$ we define\begin{eqnarray}
G_{-} & \equiv & \mathcal{D}\Phi_{+}\,,\\
G_{+} & \equiv & -\mathcal{D}\Phi_{-}\,.\end{eqnarray}
 Both $G_{-}$ and $G_{+}$ are complex, moduli-dependent fluxes analogous
to the type IIB flux $G_{3}=F_{3}-\tau H_{3}$. The superpotential
can then be written in two ways,\begin{equation}
W=\int\left\langle G_{-},\Phi_{-}\right\rangle =\int\left\langle G_{+},\Phi_{+}\right\rangle .\end{equation}
 The first expression is appropriate for analyzing the $D_{i}W=0$
minimization conditions, while the second is appropriate for analyzing
the $D_{a}W=0$ conditions. The resulting equations will be sufficiently
similar to the GKP attractor equations, that we can reuse solutions
computed in that context.

Expanding the complex fluxes on the real basis we find: \begin{eqnarray}
G_{-} & \equiv & \mathcal{D}\Phi_{+}=m_{-}^{I}\alpha_{I}-e_{I}^{-}\beta^{I}\,\end{eqnarray}
 where \begin{eqnarray}
m_{-}^{I} & \equiv & q_{A}^{\hpa I}X^{A}-q^{AI}G_{A}\,,\label{eq:m-minus}\\
e_{I}^{-} & \equiv & -q_{AI}X^{A}+q_{\hpa I}^{A}G_{A}\,.\label{eq:e-minus}\end{eqnarray}
 Similarly, for the odd complex flux, \begin{eqnarray}
G_{+} & \equiv & -\mathcal{D}\Phi_{-}=-m_{+}^{A}\omega_{A}+e_{A}^{+}\omega^{A}~,\end{eqnarray}
 where \begin{eqnarray}
m_{+}^{A} & \equiv & -\left(q_{\hpa I}^{A}Z^{I}+q^{AI}F_{I}\right)\,,\label{eq:m-plus}\\
e_{A}^{+} & \equiv & -\left(q_{AI}Z^{I}+q_{A}^{\hpa I}F_{I}\right)\,.\label{eq:e-plus}\end{eqnarray}
 These expansions differ from $G_{3}=F_{3}-\tau H_{3}$ (and generalizations),
used in other attractor analyses, in that they are not linear in the
moduli, but have more complicated dependence that enters through the
$F_{I}$ and $G_{A}.$

The tadpole constraints for these complex fluxes are interesting.
Mukai pairings between even and odd forms automatically vanish and
the antisymmetry of the Mukai pairing implies $\int\left\langle G_{\pm},G_{\pm}\right\rangle =0$, 
so the only pairings we need to consider are: \begin{eqnarray}
\int\left\langle G_{+},\overline{G}_{+}\right\rangle  & = & \left(\Phi_{-}^{T}\mathbb{Q}^{T}\right)\mathbb{S}_{+}\left(\mathbb{Q}\overline{\Phi}_{-}\right)\\
 & = & \Phi_{-}^{T}\left(\mathbb{Q}^{T}\mathbb{S}_{+}\mathbb{Q}\right)\overline{\Phi}_{-}\,,\\
\int\left\langle G_{-},\overline{G}_{-}\right\rangle  & = & \left(\Phi_{+}^{T}\mathbb{S}_{+}^{T}\mathbb{Q}\,\mathbb{S}_{-}\right)\mathbb{S}_{-}\left(\mathbb{S}_{-}^{T}\mathbb{Q}^{T}\mathbb{S}_{+}\overline{\Phi}_{+}\right)\\
 & = & \Phi_{+}^{T}\mathbb{S}_{+}^{T}\left(\mathbb{Q}\,\mathbb{S}_{-}\mathbb{Q}^{T}\right)\mathbb{S}_{+}\overline{\Phi}_{+}\,.\label{eq:G-mukai}\end{eqnarray}
 In view of the tadpole constraints \eqref{eq:Q-tadpole} for $\mathbb{Q}$
we conclude that: \begin{equation}
\int\left\langle G_{+},\overline{G}_{+}\right\rangle =\int\left\langle G_{-},\overline{G}_{-}\right\rangle =0\,.\label{GG}\end{equation}
 Recall that, in the case of the IIB O3/O7 attractors, $\int G_{3}\wedge\overline{G}_{3}$
represents a charge that needs to be cancelled, e.g. by a number of
O3 planes. Hence (\ref{GG}) states that the heterotic attractors
under consideration are source-free, as was to be expected.

Next, let us write the attractor equations for $G_{+}$. Since $G_{+}=-\mathbb{Q}\Phi_{-},$
we can simply contract \eqref{eq:Q-att} with $\Phi_{-}$ to find
an attractor equation for $G_{+}:$ \begin{equation}
G_{+}=-\mathbb{Q}\Phi_{-}=-\frac{1}{2}\left\{ \overline{C}\overline{\Phi}_{+}+\overline{L}^{\overline{AI}}\overline{\partial_{A}\Phi}_{+}\overline{K}_{I}+\widetilde{L}^{A\overline{I}}\partial_{A}\Phi_{+}\overline{K}_{I}\right\} \int\left\langle \overline{\Phi}_{-},\Phi_{-}\right\rangle .\label{eq:QPhi-}\end{equation}
 We used \eqref{eq:dI-minus} to simplify $\int\left\langle \overline{\partial_{I}\Phi}_{-},\Phi_{-}\right\rangle .$
When we impose \eqref{eq:LK-1}, this simplifies to \begin{equation}
G_{+}=\overline{C_{+}\Phi}_{+}+L_{+}^{A}\partial_{A}\Phi_{+}\,,\label{eq:Even-Att}\end{equation}
 where \begin{eqnarray}
C_{+} & \equiv & -\frac{1}{2}C\int\left\langle \Phi_{-},\overline{\Phi}_{-}\right\rangle \,,\label{eq:C+Def}\\
L_{+}^{A} & \equiv & \frac{1}{2}\widetilde{L}^{A\overline{I}}\overline{K}_{\overline{I}}\int\left\langle \Phi_{-},\overline{\Phi}_{-}\right\rangle \,.\label{eq:L+Def}\end{eqnarray}
 Similarly, the fluxes are related to $\Phi_{-}$ as \begin{equation}
G_{-}=\overline{C_{-}\Phi}_{-}+L_{-}^{I}\partial_{I}\Phi_{-}\,\label{eq:Odd-Att}\end{equation}
 with \begin{eqnarray}
C_{-} & \equiv & -\frac{1}{2}C\int\left\langle \Phi_{+},\overline{\Phi}_{+}\right\rangle \,,\label{eq:C-Def}\\
L_{-}^{I} & \equiv & \frac{1}{2}\overline{\widetilde{L}}^{\overline{A}I}\overline{K}_{\overline{A}}\int\left\langle \Phi_{+},\overline{\Phi}_{+}\right\rangle \,.\label{eq:L-Def}\end{eqnarray}
 The projected attractor equations \eqref{eq:Even-Att} and \eqref{eq:Odd-Att}
constitute only $4\left(h_{-}+1\right)+4\left(h_{+}+1\right)$ complex
equations, far fewer than the roughly $4h_{-}h_{+}$ that appear in
the full attractor equations \eqref{eq:Q-att}. However, they do determine
all of the moduli. It is just some of the components of $\widetilde{L}^{A\overline{I}}$
that must be determined subsequently, by considering the full attractor
equations.

The projected attractor equations (\ref{eq:Even-Att}) and (\ref{eq:Odd-Att})
each take the same form as the complex attractor equation for IIB O3/O7
compactifications obtained in \citep{Larsen:2009fw}. This will be
helpful when it comes to solving them, despite a difference: in the
IIB context there is a constraint from the $D_{\tau}W=0$ condition
that, translated to the present setting, would cancel the term in
$G_{+}$ proportional to $\Phi_{+}.$ In the present case there is
no analogous condition, and so no additional constraint is required.

In the preceding, we used both the full attractor equations \eqref{eq:Q-att}
and the constraints \eqref{eq:LK-1}, \eqref{eq:LK-2} to derive the
projected attractor equations \eqref{eq:Even-Att}, \eqref{eq:Odd-Att}.
In fact the projected attractor equations are simply equivalent to
the constraints, upon application of the full attractor equations.
For example, if we subtract the $G_{+}$ attractor equation \eqref{eq:Even-Att}
from the $\mathbb{Q}$ attractor equation contracted with $\Phi_{-},$
i.e. \eqref{eq:QPhi-}, the remainder is \begin{equation}
0=\frac{1}{2}\left\langle \Phi_{-},\overline{\Phi}_{-}\right\rangle \overline{L}^{\overline{AI}}\overline{\partial_{A}\Phi}_{+}\overline{K}_{I}\,.\end{equation}
 If we expand out $\overline{\partial_{A}\Phi}_{+}$ in components,
we find \begin{eqnarray}
0 & = & \overline{L}^{\overline{AI}}\overline{K}_{I}\,,\label{eq:LK-proof-1}\\
0 & = & \overline{L}^{\overline{BI}}\overline{G}_{\overline{AB}}\overline{K}_{I}\,.\label{eq:LK-proof-2}\end{eqnarray}
 The first line is just the constraint \eqref{eq:LK-1}, and once
the first line is imposed the second line is automatically satisfied.
Thus we conclude that in \eqref{eq:RealAtt-1}-\eqref{eq:LConstraint-2}
we can replace the constraints \eqref{eq:LConstraint-1}-\eqref{eq:LConstraint-2}
with the complex attractor equations \eqref{eq:Even-Att} and \eqref{eq:Odd-Att}, 
despite the fact that the latter constitute twice as many equations as 
\eqref{eq:LConstraint-1}-\eqref{eq:LConstraint-2}.

\subsection{\label{sub:SU3-AttractorEqs}Imposing $SU(3)$-Structure}

So far we have been working with a generic set of fluxes $\mathbb{Q},$
which is appropriate for $SU(3)\times SU(3)$-structure compactifications.
We now set $q^{AI}=q_{\hpa I}^{A}=0,$ which as explained in Section
\ref{sub:Fluxes}, implements the specialization to $SU(3)$-structure
compactifications. Hence the real attractor equations \eqref{eq:RealAtt-1}
and \eqref{eq:RealAtt-2} acquire the form: \begin{eqnarray}
0 & = & \mbox{Re}\left\{ CX^{A}Z^{I}+L^{AI}+\widetilde{L}^{A\overline{I}}\right\} ,\label{eq:SU3-1}\\
0 & = & \mbox{Re}\left\{ CX^{A}F_{I}+L^{AJ}F_{IJ}+\widetilde{L}^{A\overline{J}}\overline{F}_{\overline{IJ}}\right\} .\label{eq:SU3-2}\end{eqnarray}
 Note that this system of equations is linear in the $L^{AI}$. Thus,
we can solve \eqref{eq:SU3-1} by: \begin{equation}
L^{AI} =-\left(CX^{A}Z^{I}+\overline{\widetilde{L}}^{\overline{A}I}\right)+i\xi^{AI}\,,\label{eq:SU3-sol}\end{equation}
 where the $\xi^{AI}$ are some undetermined real quantities. We then
substitute this into the second set of equations and use the homogeneity
relation $Z^{J}F_{IJ}=F_{I}$ to find \begin{eqnarray}
0 & = & \xi^{AJ}\mbox{Im}\left(F_{IJ}\right)\,.\end{eqnarray}
 Therefore, \eqref{eq:SU3-sol} is a solution to \eqref{eq:SU3-1}
and \eqref{eq:SU3-2}, when $\xi^{AI}=0.$ This solution is \emph{unique},
as long as the mild regularity condition \,$\mbox{det}\left[\mbox{Im}\left(F_{IJ}\right)\right]\neq0$
\,is satisfied.\footnote{Interestingly, our considerations here imply, in particular, that 
the attractors of \citep{DallAgata:2006nr,Anguelova:2008fm} are empty unless 
$\mbox{det}\left[\mbox{Im}\left(F_{IJ}\right)\right] = 0$.} We can substitute 
\eqref{eq:SU3-sol} back into our
previous attractor equations to see how they simplify. We start with
the expression for $\mathbb{Q}$ \eqref{eq:Q-att}, which becomes\begin{eqnarray}
\mathbb{Q} & = & \mbox{Re}\left\{ - \overline{\widetilde{L}}^{\overline{A}I} \Phi_{+}\left(\partial_{I}\Phi_{-}\right)^{T}+ \widetilde{L}^{A\overline{I}} \partial_{A}\Phi_{+}\left(\overline{\partial_{I}\Phi}_{-}\right)^{T}\right\} \mathbb{S}_{-}\\
 & = & -2\mbox{Im}\left\{ \partial_{A}\Phi_{+}\right\} \mbox{Im}\left\{ \widetilde{L}^{A\overline{I}} \overline{\partial_{I}\Phi}_{-}^{T}\right\} \mathbb{S}_{-}\,.\end{eqnarray}
 Expanded in components, this becomes: \begin{eqnarray}
q_{A}^{\hpa I} & = & 2\mbox{Im}\left\{ G_{AB}\right\} \mbox{Im}\left\{ \widetilde{L}^{B\overline{I}}\right\} ,\label{eq:RealAtt-SU3-1}\\
q_{AI} & = & -2\mbox{Im}\left\{ G_{AB}\right\} \mbox{Im}\left\{ \widetilde{L}^{B\overline{J}}\overline{F}_{\overline{IJ}}\right\} \\
 & = & -q_{A}^{\hpa I}\mbox{Re}\left\{ F_{IJ}\right\} +2\mbox{Im}\left\{ G_{AB}\right\} \mbox{Re}\left\{ \widetilde{L}^{B\overline{J}}\right\} \mbox{Im}\left\{ F_{IJ}\right\} .\label{eq:RealAtt-SU3-2}\end{eqnarray}
 We could also have derived these by substituting \eqref{eq:SU3-sol}
into \eqref{eq:RealAtt-3} and \eqref{eq:RealAtt-4}.

We will also need the projected attractor equations after $SU(3)$-structure
is imposed. Setting $q_{\hpa I}^{A}=q^{AI}=0$ means that the complex
fluxes $m_{+}^{A}$ also vanish. Expanding \eqref{eq:Even-Att}
on the basis forms $\left\{ \omega_{A},\omega^{A}\right\} $, we find:\begin{eqnarray}
m_{+}^{A}=0 & = & \overline{C_{+}X}^{A}+L_{+}^{A}\,,\\
e_{A}^{+} & = & \overline{C_{+}G}_{A}+L_{+}^{B}G_{AB}\,.\end{eqnarray}
 Clearly, we can solve immediately the magnetic attractor equation.
The electric one then simplifies to: \begin{eqnarray}
e_{A}^{+} & = & \overline{C_{+}G}_{A}-\overline{C_{+}X}^{B}G_{AB}\\
 & = & -2i\overline{C_{+}X}^{B}\mbox{Im}\left(G_{AB}\right),\label{eq:Complex-SU3-Att}\end{eqnarray}
 where we have used the homogeneity relation $X^{B}G_{AB}=G_{A}$
to obtain the last line. The projected attractor equations for $e_I^-$ and $m^I_-$ are formally unchanged,
\begin{eqnarray*}
m_{-}^{I} & = & \overline{C_{-}Z}^{I}+L_{-}^{I}\,,\\
e_{I}^{-} & = & \overline{C_{-}F}_{I}+L_{-}^{J}F_{IJ}\,.\end{eqnarray*}

We can now spell out the strategy we propose for solving the heterotic
attractor equations. First, solve the complex attractor equations
\eqref{eq:Even-Att} and \eqref{eq:Odd-Att} to determine the moduli
$x^{a}$ and $z^{i},$ as well as the mass parameter $CX^{0}Z^{0}.$
These completely determine $F_{IJ}$ and $G_{AB},$ so it is then
a matter of linear algebra to solve \eqref{eq:RealAtt-SU3-1} and
\eqref{eq:RealAtt-SU3-2} for $\widetilde{L}^{A\overline{I}}$. The 
remaining mass parameters, $L^{AI}$, are then simply related to the 
$\widetilde{L}^{A\overline{I}}$ by \eqref{eq:SU3-sol}.

\section{\label{sec:Generating-Functions}Generating Functions}

In \eqref{eq:Even-Att} and \eqref{eq:Odd-Att} we have essentially
found two copies of the IIB O3/O7 attractor equations, as presented
e.g. in \citep{Larsen:2009fw}. With this in mind, we will demonstrate
that associated to each set of attractor equations is a generating
function. Solutions of the two sets of attractor equations can then
be written as derivatives of the appropriate generating function.
When we compare the flux attractor equations to the black hole 
ones, the generating functions here play the same role as the black
hole entropy and thus are quantities of great interest.

We will develop these generating functions in two steps. First, we
will introduce two {}``bare'' generating functions whose derivatives
reproduce the conditions $D_{i}W=0$ (written as holomorphic functions
of the $e_{I}^{-}$ and $m_{-}^{I}$) and $D_{a}W=0$ (written as
holomorphic functions of the $e_{A}^{+}$ and $m_{+}^{A}$). This
will give us two generating functions, one that depends only on $e_{I}^{-}$
and $m_{-}^{I}$, and one that depends only on $e_{A}^{+}$ and $m_{+}^{A}.$
We will then {}``dress'' each generating function so that it reproduces
the complementary set of partial minimization conditions. A similar
dressing is required for the IIB O3/O7 generating function in order
to properly reproduce the $D_{\tau}W=0$ condition.

\subsection{Bare Generating Functions}

We begin by writing out the complex attractor equation for the odd
moduli \eqref{eq:Odd-Att} in components:\begin{eqnarray}
m_{-}^{I} & = & \overline{C_{-}Z}^{I}+L_{-}^{I}\,,\label{eq:Odd-Mag}\\
e_{I}^{-} & = & \overline{C_{-}F}_{I}+L_{-}^{J}F_{IJ}\,.\label{eq:Odd-Elec}\end{eqnarray}
 In these equations, we consider $C_{-}Z^{I}$ and $L_{-}^{I}$ as
the independent variables, while $C_{-}F_{I}$ and $L_{-}^{J}F_{IJ}$
are their symplectic partners. We can formally solve
the magnetic attractor equation (\ref{eq:Odd-Mag}) by the change
of variables:\begin{eqnarray}
C_{-}Z^{I} & = & \frac{1}{2}\left(\overline{m}_{-}^{I}-i\overline{\phi}_{-}^{I}\right),\label{eq:CZ-minus}\\
L_{-}^{I} & = & \frac{1}{2}\left(m_{-}^{I}-i\phi_{-}^{I}\right),\label{eq:L-minus}\end{eqnarray}
 where we have introduced complex potentials $\phi_{-}^{I}$. Then we 
just need to solve the electric attractor equations \eqref{eq:Odd-Elec}
for the potentials $\phi_{-}^{I}.$ It is also useful to introduce
the dual potentials $\theta_{I}^{-}$ through \begin{eqnarray}
C_{-}F_{I} & = & \frac{1}{2}\left(\overline{e}_{I}^{-}-i\overline{\theta}_{I}^{-}\right),\label{eq:CF-minus}\\
L_{-}^{J}F_{IJ} & = & \frac{1}{2}\left(e_{I}^{-}-i\theta_{I}^{-}\right)~.\label{eq:LF-minus}\end{eqnarray}
 These are the symplectic partners of $\phi_{-}^{I}$.

Relating derivatives with respect to $C_{-}Z^{I}$ and $L^{I}$ to
derivatives with respect to the fluxes and potentials: \begin{eqnarray}
\frac{\partial}{\partial\overline{C_{-}Z}^{I}} & = & \frac{\partial}{\partial m_{-}^{I}}-i\frac{\partial}{\partial\phi_{-}^{I}}\,,\\
\frac{\partial}{\partial L_{-}^{I}} & = & \frac{\partial}{\partial m_{-}^{I}}+i\frac{\partial}{\partial\phi_{-}^{I}}\,,\end{eqnarray}
 the electric attractor equation simplifies: \begin{eqnarray}
e_{I}^{-} & = & -\frac{\partial}{\partial\overline{\phi}_{-}^{I}}4\mbox{Im}\left(C_{-}F_{K}L_{-}^{K}\right).\label{eq:elattra}\end{eqnarray}
 The expression $4\mbox{Im}\left(C_{-}F_{K}L_{-}^{K}\right)$ is treated
as a function of $m_{-}^{I}$ and $\phi_{-}^{I}.$ Thus it is defined
for a mixed ensemble, so its analogue in black hole attractors is
the free energy of the black hole. Since the electric attractor equation
(\ref{eq:elattra}) has been written in the form of a thermodynamic
relation, we can formally solve it with the simple Legendre transform:
\begin{eqnarray}
\mathfrak{g}_{-}\left(e_{I}^{-},m_{-}^{I}\right) & = & 4\mbox{Im}\left(C_{-}F_{K}L_{-}^{K}\right)+e_{K}^{-}\overline{\phi}_{-}^{K}+\overline{e}_{K}^{-}\phi_{-}^{K}\,.\label{eq:g-minus-long}\end{eqnarray}
 We refer to $\mathfrak{g}_{-}$ as the bare generating function.
The factors of $i$ in (\ref{eq:CZ-minus})-(\ref{eq:L-minus}) were
introduced to ensure that this function is real. The bare generating
function is defined in an ensemble of fixed fluxes, and so is analogous
to the black hole entropy. The transform of the electric attractor
equation is the formal solution of the attractor equations, \begin{equation}
\phi_{-}^{I}=\frac{\partial\mathfrak{g}_{-}}{\partial\overline{e}_{I}^{-}}\,,\label{eq:phi-minus-1}\end{equation}
 while symplectic covariance implies \begin{equation}
\theta_{I}^{-}=-\frac{\partial\mathfrak{g}_{-}}{\partial\overline{m}_{-}^{I}}\,.\label{eq:theta-minus}\end{equation}

Clearly, the {}``$-$'' subscripts were not an important part of
the preceding considerations. If we allow a full set of $SU(3)\times SU(3)$
fluxes, we can introduce a complementary set of potentials for the
Kähler moduli, \begin{eqnarray}
C_{+}X^{A} & = & \frac{1}{2}\left(\overline{m}_{+}^{A}-i\overline{\phi}_{+}^{A}\right), \label{eq:CX}\\
C_{+}G_{A} & = & \frac{1}{2}\left(\overline{e}_{A}^{+}-i\overline{\theta}_{A}^{+}\right),\\
L_{+}^{A} & = & \frac{1}{2}\left(m_{+}^{A}-i\phi_{+}^{A}\right),\\
L_{+}^{B}G_{AB} & = & \frac{1}{2}\left(e_{A}^{+}-i\theta_{A}^{+}\right), \label{eq:LG}\end{eqnarray}
 and a generating function\begin{equation}
\mathfrak{g}_{+}\left(e_{A}^{+},m_{+}^{A}\right)=4\mbox{Im}\left(C_{+}G_{B}L_{+}^{B}\right)+e_{B}^{+}\overline{\phi}_{+}^{B}+\overline{e}_{B}^{+}\phi_{+}^{B}\,.\label{eq:g-plus-long}\end{equation}
 The expressions for the potentials are similar to those given above:
\begin{eqnarray}
\phi_{+}^{A} & = & \frac{\partial\mathfrak{g}_{+}}{\partial\overline{e}_{A}^{+}}\,,\label{eq:phi-plus}\\
\theta_{A}^{+} & = & -\frac{\partial\mathfrak{g}_{+}}{\partial\overline{m}_{+}^{A}}\,.\label{eq:theta-plus}\end{eqnarray}

We explained in Section \ref{sub:Complex-Attractor-Equations} that 
one specializes to $SU(3)$-structure by setting the $m_{+}^{A}$
to zero.  While this prescription is correct for the moduli and mass parameters, it must be modified slightly for the generating function.  Because the $\theta_A^+$ remain non-vanishing even after we set $m^A_+=0$, we must retain terms linear in the $m^A_+$ in $\mathfrak{g}_+$ in order to correctly reproduce \eqref{eq:theta-plus}.

\subsection{$\mathfrak{g}_{-}$ and $\mathfrak{g}_{+}$ at the Attractor Point}

Before dressing $\mathfrak{g}_{-}$ and $\mathfrak{g}_{+}$, we derive
two useful expressions for their values at the attractor point(s).
For concreteness we will work with the complex structure moduli, but
the expressions will apply just as well for the Kähler moduli.

We begin by writing two equivalent expressions for $4\mbox{Im}\left(C_{-}F_{K}L_{-}^{K}\right),$
the Legendre transform of $\mathfrak{g}_{-}$. First we use \eqref{eq:CF-minus}
and \eqref{eq:L-minus} to find: \begin{eqnarray}
4\mbox{Im}\left(C_{-}F_{K}L_{-}^{K}\right) & = & \mbox{Im}\left[\left(\overline{e}_{K}^{-}-i\overline{\theta}_{K}^{-}\right)\left(m_{-}^{K}-i\phi_{-}^{K}\right)\right]\\
 & = & \mbox{Im}\left[\overline{e}_{K}^{-}m_{-}^{K}-i\overline{\theta}_{K}^{-}m_{-}^{K}-i\overline{e}_{K}^{-}\phi_{-}^{K}-\overline{\theta}_{K}^{-}\phi_{-}^{K}\right].\label{eq:CFL1}\end{eqnarray}
 Next, the homogeneity of $F_{I}$ implies that $\mbox{Im}\left(C_{-}F_{K}L_{-}^{K}\right)=\mbox{Im}\left(C_{-}Z^{J}F_{JK}L_{-}^{K}\right).$
If we substitute \eqref{eq:CZ-minus} and \eqref{eq:LF-minus} into
the last equation, we find: \begin{eqnarray}
4\mbox{Im}\left(C_{-}F_{K}L_{-}^{K}\right) & = & \mbox{Im}\left[\left(e_{K}^{-}-i\theta_{K}^{-}\right)\left(\overline{m}_{-}^{K}-i\overline{\phi}_{-}^{K}\right)\right]\\
 & = & \mbox{Im}\left[e_{K}^{-}\overline{m}_{-}^{K}-i\theta_{K}^{-}\overline{m}_{-}^{K}-ie_{K}^{-}\overline{\phi}_{-}^{K}-\theta_{K}^{-}\overline{\phi}_{-}^{K}\right].\label{eq:CFL2}\end{eqnarray}
 Comparing \eqref{eq:CFL1} and \eqref{eq:CFL2}, we conclude that
\begin{equation}
\mbox{Im}\left[\overline{e}_{K}^{-}m_{-}^{K}-\overline{\theta}_{K}^{-}\phi_{-}^{K}\right]=0\end{equation}
 and \begin{equation}
4\mbox{Im}\left(C_{-}F_{K}L_{-}^{K}\right)=-\mbox{Re}\left[\theta_{K}^{-}\overline{m}_{-}^{K}+\phi_{-}^{K}\overline{e}_{K}^{-}\right].\label{eq:Nice-CFL}\end{equation}
 Referring to \eqref{eq:g-minus-long}, we conclude that \begin{eqnarray}
\mathfrak{g}_{-} & = & -\mbox{Re}\left[\overline{\theta}_{K}^{-}m_{-}^{K}+\overline{e}_{K}^{-}\phi_{-}^{K}\right]+2\mbox{Re}\left[\overline{e}_{I}^{-}\phi_{-}^{I}\right]\\
 & = & \mbox{Re}\left[e_{K}^{-}\overline{\phi}_{-}^{K}-m_{-}^{K}\overline{\theta}_{K}^{-}\right].\label{eq:nice-gfrak-minus}\end{eqnarray}
 An analogous computation gives \begin{equation}
\mathfrak{g}_{+}=\mbox{Re}\left[e_{A}^{+}\overline{\phi}_{+}^{A}-m_{+}^{A}\overline{\theta}_{A}^{+}\right].\label{eq:nice-gfrak-plus}\end{equation}

One reason these expressions are useful is that they are easily compared
with the attractor value of the superpotential. To see how this comes
about, we compute: \begin{eqnarray}
W=\int\left\langle G_{-},\Phi_{-}\right\rangle =\overline{C}_{-}\int\left\langle \overline{\Phi}_{-},\Phi_{-}\right\rangle =ie^{-K_{-}}\overline{C}_{-}\,\,.\label{eq:WC-Complex}\end{eqnarray}
 We used the projected attractor equation \eqref{eq:Odd-Att} and
the expression \eqref{eq:K} for $K_{-}$. This implies the following
chain of equalities: \begin{equation}
e^{K_{-}}\left|W\right|^{2}=-iC_{-}W=e^{-K_{-}}\left|C_{-}\right|^{2}\,.\label{eq:TripleEquality}\end{equation}
 Since the expressions at the ends are manifestly real, we conclude
that $C_{-}W$ (evaluated at the attractor point) is purely imaginary.
We can compute an expression for $C_{-}W$ analogous to \eqref{eq:Nice-CFL}
and \eqref{eq:nice-gfrak-minus}:\begin{eqnarray}
C_{-}W & = & \int\left\langle G_{-},C_{-}\Phi_{-}\right\rangle \\
 & = & C_{-}\left(Z^{I}e_{I}^{-}-F_{I}m_{-}^{I}\right)\\
 & = & \frac{1}{2}\left\{ \left(\overline{m}_{-}^{I}e_{I}^{-}-\overline{e}_{I}^{-}m_{-}^{I}\right)-i\left(\overline{\phi}_{-}^{I}e_{I}^{-}-\overline{\theta}_{I}^{-}m_{-}^{I}\right)\right\} \\
 & = & \frac{1}{2}\left\{ \int\left\langle G_{-},\overline{G}_{-}\right\rangle -i\left(\overline{\phi}_{-}^{I}e_{I}^{-}-\overline{\theta}_{I}^{-}m_{-}^{I}\right)\right\} \end{eqnarray}
 We demonstrated above that $C_{-}W$ must be purely imaginary. Since
the Mukai pairing is antisymmetric, $\int\left\langle G_{-},\overline{G}_{-}\right\rangle $
is also purely imaginary. Hence $\overline{\phi}_{-}^{I}e_{I}^{-}-\overline{\theta}_{I}^{-}m_{-}^{I}$
must be purely real (at the attractor point) and, in view of (\ref{eq:nice-gfrak-minus}),
equal to $\mathfrak{g}_{-}$. This implies a relationship at the attractor
point%
\footnote{This is the analogue of eq. (5.33) in \citep{Larsen:2009fw}, although
there the generating function is already dressed and so the normalization
of the $\int G_{3}\wedge\overline{G}_{3}$ term is different.%
}: \begin{eqnarray}
\mathfrak{g}_{-} & = & 2iC_{-}W-i\int\left\langle G_{-},\overline{G}_{-}\right\rangle .\label{eq:g-on-shell-minus}\end{eqnarray}
 For the even moduli we similarly find: \begin{equation}
\mathfrak{g}_{+}=2iC_{+}W-i\int\left\langle G_{+},\overline{G}_{+}\right\rangle .\label{eq:g-on-shell-plus}\end{equation}
 Note that when we use the standard embedding for the 10D gauge field,
the tadpole constraints require that $\int\left\langle G_{-},\overline{G}_{-}\right\rangle =\int\left\langle G_{+},\overline{G}_{+}\right\rangle =0.$

It would be overly conservative to apply the above expressions only
at the attractor point, since in each case we have applied only a
subset of the attractor equations. For example, in deriving the expression
for $\mathfrak{g}_{-}$ we used the $D_{i}W=0$ conditions to eliminate
the dependence on the $z^{i},$ but $\mathfrak{g}_{-}$ retains (through
$e_{I}^{-}$ and $m_{-}^{I}$) some dependence on the $x^{a}$. We
now verify that \eqref{eq:g-on-shell-minus} holds throughout the
even moduli space, and that \eqref{eq:g-on-shell-plus} holds throughout
the odd moduli space.

We begin by computing the derivative of $\mathfrak{g}_{-}$ with respect
to the $X^{A},$ while holding the \emph{real }fluxes fixed: \begin{eqnarray}
\left.\frac{\partial\mathfrak{g}_{-}}{\partial X^{A}}\right|_{\mathbb{R}} & = & \frac{\partial\mathfrak{g}_{-}}{\partial e_{I}^{-}}\frac{\partial e_{I}^{-}}{\partial X^{A}}+\frac{\partial\mathfrak{g}_{-}}{\partial m_{-}^{I}}\frac{\partial m_{-}^{I}}{\partial X^{A}}\\
 & = & \overline{\phi}_{-}^{I}\frac{\partial e_{I}^{-}}{\partial X^{A}}-\overline{\theta}_{I}^{-}\frac{\partial m_{-}^{I}}{\partial X^{A}}\,\,.\end{eqnarray}
 If we rearrange \eqref{eq:CZ-minus} and \eqref{eq:CF-minus} to
find nice expressions for the potentials, \begin{eqnarray}
\overline{\phi}_{-}^{I} & = & 2iC_{-}Z^{I}-i\overline{m}_{-}^{I}\,,\\
\overline{\theta}_{I}^{-} & = & 2iC_{-}F_{I}-i\overline{e}_{I}^{-}\,,\end{eqnarray}
 then the $X^{A}$-derivatives become:\begin{eqnarray}
\left.\frac{\partial\mathfrak{g}_{-}}{\partial X^{A}}\right|_{\mathbb{R}} & = & 2iC_{-}\left(Z^{I}\frac{\partial e_{I}^{-}}{\partial X^{A}}-F_{I}\frac{\partial m_{-}^{I}}{\partial X^{A}}\right)-i\left(\overline{m}_{-}^{I}\frac{\partial e_{I}^{-}}{\partial X^{A}}-\overline{e}_{I}^{-}\frac{\partial m_{-}^{I}}{\partial X^{A}}\right)\\
 & = & 2iC_{-}\int\left\langle \frac{\partial G_{-}}{\partial X^{A}},\Phi_{-}\right\rangle -i\int\left\langle \frac{\partial G_{-}}{\partial X^{A}},\overline{G}_{-}\right\rangle \\
 & = & \frac{\partial}{\partial X^{A}}\left(2iC_{-}W-i\int\left\langle G_{-},\overline{G}_{-}\right\rangle \right)~,\label{eq:nice-dAg-minus}\end{eqnarray}
 where we have used the holomorphy of $G_{-}$ in the $X^{A}$ to
reassemble the superpotential and $\int\left\langle G_{-},\overline{G}_{-}\right\rangle .$
We have thus verified that $X^{A}$-derivatives of the full $\mathfrak{g}_{-}$
are equivalent to $X^{A}$-derivatives of \eqref{eq:g-on-shell-minus}.
As usual, an analogous result holds for $\mathfrak{g}_{+}.$

\subsection{\label{sub:Dressing}The Dressed Generating Functions}

The bare generating function $\mathfrak{g}_{-}$ was constructed so
derivatives with respect to the complex fluxes $e_{I}^{-}$ and $m_{-}^{I}$
give the desired potentials. The derivatives with respect to $X^{A}$
have no analogous interpretation \textit{a priori} but we will show
next that the {}``dressed'' generating function
\begin{equation}
\mathcal{G}_{-}=f(X^{A})\mathfrak{g}_{-}~,\end{equation}
 for a suitable dressing factor $f(X^{A})$ has derivatives: \begin{equation}
\left.\frac{\partial\mathcal{G}_{-}}{\partial x^{a}}\right|_{\mathbb{R}}\propto D_{a}W\,.\label{eq:g-deri}\end{equation}
 Thus the F-flatness conditions $D_{a}W=0$ will be satisfied by extremizing
$\mathcal{G}_{-}$ with respect to the $x^{a}.$ This construction
is motivated by the IIB O3/O7 generating function, where the derivative
with respect to $\tau$ (which is analogous to the $X^{A},$ in that
they both appear in the definition of the respective complex fluxes)
is proportional to $D_{\tau}W$.

The dressing factor is in fact largely determined by the symmetries
of the problem. Heterotic attractors are built on two symplectic sections,
$\Phi_{+}$ and $\Phi_{-},$ and because of this have two separate
rescaling symmetries:\begin{eqnarray}
\Phi_{-} & \to & \lambda_{-}(z^{i})\Phi_{-}\,,\\
\Phi_{+} & \to & \lambda_{+}(x^{a})\Phi_{+}\,,\end{eqnarray}
 where the $\lambda_{\pm}$ are (nowhere-vanishing) holomorphic functions
of the $z^{i}$ and $x^{a}.$ Objects in the theory generally transform
nontrivially under these rescalings but physical quantities must be
invariant. Considering for example the attractor equation \eqref{eq:Odd-Att},
the objects $G_{-},$ $C_-\Phi_{-},$ $\partial_{I}\Phi_{-},$ are all
invariant under rescalings of $\Phi_{-}$ (this follows from the fact that 
$\mathbb{Q}$ has to be invariant); but, as one can see from
(\ref{eq:C-Def}-\ref{eq:L-Def}), they transform as $(1,1)$ tensors
under rescalings of $\Phi_{+}.$ The bare generating function $\mathfrak{g}_{-}$
is then a $(1,1)$ tensor under rescalings of $\Phi_{+}$ as well.
The simplest way to form an invariant employs the Kähler potential
and thus introduces the dressed generating function as \begin{equation}
\mathcal{G}_{-}=e^{K_{+}}\mathfrak{g}_{-}\,.\label{eq:G-minus}\end{equation}
 We now verify that this dressing factor gives the desired derivative
(\ref{eq:g-deri}).

It is advantageous at this point to make the change of variables $\left\{ X^{A}\right\} \to\left\{ X^{0},x^{a}\right\} .$
Since $\mathcal{G}_{-}$ is invariant under rescalings of $\Phi_{+},$
the $X^{0}$-derivative of $\mathcal{G}_{-}$ vanishes, while \eqref{eq:nice-dAg-minus}
and \eqref{eq:G-minus} give \begin{eqnarray}
\left.\frac{\partial\mathcal{G}_{-}}{\partial x^{a}}\right|_{\mathbb{R}} & = & e^{K_{+}}\left[\frac{\partial\mathfrak{g}_{-}}{\partial x^{a}}+\mathfrak{g}_{-}\frac{\partial K_{+}}{\partial x^{a}}\right]\\
 & = & e^{K_{+}}\left[2iC_{-}D_{a}W-i\int\left\langle D_{a}G_{-},\overline{G}_{-}\right\rangle \right]~.\label{eq:G-derint}\end{eqnarray}
The final term vanishes due to tadpole constraints, when we
use the standard embedding of the 10D gauge field, i.e. when $\mbox{tr}\left(R\wedge R\right)=\mbox{tr}\left(F\wedge F\right).$
This follows by taking derivative of (\ref{eq:G-mukai}) \begin{equation}
D_{a}\int\left\langle G_{-},\overline{G}_{-}\right\rangle =\left[D_{a}\Phi_{+}^{T}\right]\mathbb{S}_{+}^{T}\left(\mathbb{Q}\,\mathbb{S}_{-}\mathbb{Q}^{T}\right)\mathbb{S}_+\overline{\Phi}_{+}\,,\end{equation}
 and recalling that the tadpole constraint requires that the quantity
in round brackets vanish. The derivative (\ref{eq:G-derint}) thus
takes the claimed form (\ref{eq:g-deri}). Of course, we have relied
quite heavily on the standard embedding of the gauge field; it
would be interesting to try to construct the analogue of \eqref{eq:G-minus}
for non-standard embeddings.

The analogue of \eqref{eq:G-minus} for the even moduli is: \begin{equation}
\mathcal{G}_{+}=e^{K_{-}}\mathfrak{g}_{+}\,.\label{eq:G-plus}\end{equation}
 One can easily see that at the attractor point $\mathcal{G}_{+}$
and $\mathcal{G}_{-}$ take the same value:\begin{equation}
\mathcal{G}_{-}=\mathcal{G}_{+}=2e^{K_{-}+K_{+}}\left|W\right|^{2}.\label{eq:G-gravitino}\end{equation}
 Indeed, this follows from application of \eqref{eq:TripleEquality}
to the on-shell expressions for $\mathfrak{g}_{-}$ and $\mathfrak{g}_{+}$,
namely \eqref{eq:g-on-shell-minus} and \eqref{eq:g-on-shell-plus}
respectively. In other words, the on-shell generating functions $\mathcal{G}_{\pm}$
are both just the gravitino mass, up to a factor of the dilaton.

In summary, we have introduced two generating functions \eqref{eq:g-minus-long} (resp.
\eqref{eq:g-plus-long}) that control the complex structure moduli and  Kähler moduli, respectively.
Upon taking derivatives \eqref{eq:phi-minus-1} (resp. \eqref{eq:phi-plus}) and insertion in \eqref{eq:CZ-minus}
(resp. \eqref{eq:CX}), they yield the VEVs and some of the mass parameters.
After being dressed as in \eqref{eq:G-minus} and \eqref{eq:G-plus}, each of these also stabilize the complementary moduli, through simple
extremization and application of (\ref{eq:g-deri}).

\section{\label{sec:STU-Model}A Two-Modulus Attractor}

We now solve the heterotic flux attractor equations in a simple example.
We will consider the case of two complex moduli, described by the following 
prepotentials: \begin{eqnarray}
F & = & \frac{\left(Z\right)^{3}}{Z^{0}}\,,\label{eq:F-ex}\\
G & = & \frac{\left(X\right)^{3}}{X^{0}}\,.\label{eq:G-ex}\end{eqnarray}
 This describes the large volume/complex structure limit of any two
modulus geometry, including orbifolds of $T^{6}.$ To avoid confusion,
we will replace $i$ indices with a $z$ and $a$ indices with an
$x.$ We will be able to determine the properties of the stabilized Kähler moduli, including the bare generating function $\mathfrak{g}_{+},$
in terms of generic complex fluxes $e_A^+$. The attractor equations for the complex structure moduli space are more complicated, and we will not find explicit solutions for them as functions of generic complex fluxes.

We will be able to find explicit expressions for both $x$ and $z$ as functions of the \emph{real} fluxes by setting half of them to zero, so that $m_{-}^{0},$ $e_{z}^{-},$
and $e_{x}^{+}$ are purely real, while $m_{-}^{z},$ $e_{0}^{-},$
and $e_{0}^{+}$ are purely imaginary. Specifically, we keep only
$q_{0}^{\hpa0},$ $q_{x0},$ $q_{0z},$ and $q_{x}^{\hpa z}$. This yields a superpotential
\begin{equation}
  W=X^0 Z^0 ( q_{0}^{\hpa0} z^3-q_{x0}x-q_{0z}z-3 q_{x}^{\hpa z} x z^2),
\end{equation}
where we use the usual physical moduli $z\equiv Z/Z^0$ and $x\equiv X/X^0$. Our choice of fluxes is
sufficient to stabilize both complex moduli at purely imaginary values,
\begin{eqnarray}
z & = & iy\,,\\
x & = & it\,.\end{eqnarray}
 The fluxes must satisfy a single tadpole constraint,\begin{equation}
q_{0}^{\hpa0}q_{x0}-q_{x}^{\hpa z}q_{0z}=0\,.\label{eq:Tadpole-ex}\end{equation}
 Having specialized to this reduced set of fluxes, we will compute explicitly
the VEVs of both moduli and the mass parameters $CX^{0}Z^{0}$, 
$\widetilde{L}^{A\overline{I}}$, as functions of the real fluxes.

\subsection{Projected Attractor Equations and Generating Functions}

We begin by looking at the attractor equations for the Kähler moduli.
This requires that we compute $G_{AB}$: \begin{eqnarray}
G_{00} & = & 2x^{3}\,,\\
G_{0x} & = & -3x^{2}\,,\\
G_{xx} & = & 6x\,.\end{eqnarray}
 Substituting these into \eqref{eq:Complex-SU3-Att}, we find \begin{eqnarray}
e_{0}^{+} & = & -\overline{C_{+}X}^{0}\left(x-\overline{x}\right)^{2}\left(2x+\overline{x}\right)\,,\\
e_{x}^{+} & = & 3\overline{C_{+}X}^{0}\left(x-\overline{x}\right)^{2}.\end{eqnarray}
 These equations uniquely determine the modulus $x$ and the mass parameter
$C_{+}X^{0}$ as functions of the complex fluxes:
 \begin{eqnarray}
x & = & \frac{C_{+}X}{C_{+}X^{0}} = \frac{\overline{e}_{0}^{+}e_{x}^{+}-2e_{0}^{+}\overline{e}_{x}^{+}}{\left|e_{x}^{+}\right|^{2}}\,,\label{eq:x-ex}\\
C_{+}X^{0} & = & \overline{e}_{x}^{+}\frac{\left|e_{x}^{+}\right|^{4}}{27\left(\overline{e}_{0}^{+}e_{x}^{+}-e_{0}^{+}\overline{e}_{x}^{+}\right)^{2}}\,.\label{eq:CX0-ex}
\end{eqnarray}
 We remind the reader that the reduction to $SU(3)$-structure led
us to identify $L_{+}^{A}=-\overline{C_{+}X}^{A}$. So \eqref{eq:x-ex}
and \eqref{eq:CX0-ex} together also determine the mass parameters
from the projected attractor equations \eqref{eq:Even-Att}. Note also that
$x$ in (\ref{eq:x-ex}) is indeed purely imaginary, as claimed above, since 
$e_0^+ \overline{e}_x^+ = - \overline{e}_0^+ e_x^+$ due to $e_0^+$ being purely 
imaginary and $e_x^+$ purely real.

We will now construct the bare generating function $\mathfrak{g}_{+}$
using \eqref{eq:g-plus-long}. This requires that we compute the potentials
$\phi_{+}^{A}.$ This is quite straightforward for the Kähler moduli,
since $SU(3)$-structure implies that \begin{equation}
\phi_{+}^{A}=-2i\overline{C_{+}X}^{A}\,.\end{equation}
We also need to compute 
\begin{eqnarray}
\mbox{Im}\left(C_{+}G_{B}L_{+}^{B}\right) & = & -\mbox{Im}\left(C_{+}G_{B}\overline{C_{+}X}^{B}\right)\\
 & = & \frac{1}{2i}\left|C_{+}X^{0}\right|^{2}\left(x-\overline{x}\right)^{3}.
 \end{eqnarray}
After substituting these expressions and our solutions \eqref{eq:x-ex}
and \eqref{eq:CX0-ex} into \eqref{eq:g-plus-long}, we find: 
\begin{equation}
\mathfrak{g}_{+}=\frac{2i}{27}\frac{\left|e_{x}^{+}\right|^{4}}{\overline{e}_{0}^{+}e_{x}^{+}-e_{0}^{+}\overline{e}_{x}^{+}} 
+\mathcal{O}\left(m_{+}^{A}\right)\,.\end{equation}
While we have not computed the $\mathcal{O}\left(m_{+}^{A}\right)$
terms here, they are required in order to correctly reproduce the
dual potentials $\theta_{A}^{+}.$ 
It is straightforward to verify
that differentiating $\mathfrak{g}_{+}$ with respect to the complex
fluxes returns the potentials $\phi_{+}^{A},$ as in \eqref{eq:phi-plus}.

The projected attractor equations for the complex
structure moduli are quite complicated, as one sees from the F-flatness condition, \begin{equation}
D_{z}W\propto e_{0}^{-}+\frac{i}{3}e_{z}^{-}\left(2z-\overline{z}\right)-m_{-}^{z}z\left(z+2\overline{z}\right)+m_{-}^{0}z^{2}\overline{z}\,.\end{equation}
 This is essentially a generic cubic equation, and so we do not expect
its solutions to be particularly illuminating. We will therefore not solve for the complex structure moduli as functions of the complex fluxes $e_I^-$ and $m^I_-$.  In the next section,
we will show that the solutions for $z$ and $x$ in terms of the
\emph{real} fluxes are quite compact.

\subsection{Solutions for the Moduli}

We now solve explicitly for the moduli $z=iy$ and $x=it$ in terms
of the \emph{real} fluxes. So far we have used the complex fluxes,
which are defined in terms of the moduli and the real fluxes as:

\begin{eqnarray}
\frac{m_{-}^{0}}{X^{0}} & = & q_{0}^{\hpa0}\,,\\
\frac{m_{-}^{z}}{X^{0}} & = & iq_{x}^{\hpa z}t\,,\\
\frac{e_{0}^{-}}{X^{0}} & = & -iq_{x0}t\,,\\
\frac{e_{z}^{-}}{X^{0}} & = & -q_{0z}\,,\end{eqnarray}
 and\begin{eqnarray}
\frac{e_{0}^{+}}{Z^{0}} & = & -iy\left(q_{0z}+q_{0}^{\hpa0}y^{2}\right),\label{eq:eo-ex}\\
\frac{e_{x}^{+}}{Z^{0}} & = & -\left(q_{x0}-3q_{x}^{\hpa z}y^{2}\right).\label{eq:ex-ex}\end{eqnarray}
 If we substitute these into the F-flatness conditions for $x$ and
$z,$ we find\begin{eqnarray}
0 & = & \left(-iq_{x0}t\right)+\frac{i}{3}\left(-q_{0z}\right)y-\left(iq_{x}^{\hpa z}t\right)y^{2}+i\left(q_{0}^{\hpa0}\right)y^{3}\,,\\
0 & = & iy\left(q_{0z}+q_{0}^{\hpa0}y^{2}\right)+\frac{i}{3}\left(q_{x0}-3q_{x}^{\hpa z}y^{2}\right)t\,.\end{eqnarray}
 A bit of algebra reduces these to \begin{eqnarray}
t & = & -\frac{q_{0z}}{q_{x0}}y\,,\label{eq:ty-ex}\\
0 & = & 2q_{0z}+3q_{x}^{\hpa z}\frac{q_{0z}}{q_{x0}}y^{2}+q_{0}^{\hpa0}y^{2}\,.\end{eqnarray}

We pause to consider the branch structure of these solutions. Important
here are the so-called {}``Kähler cone constraints.'' In their simplest form, they require that the The Kähler
potentials  \begin{eqnarray}
K_{-} & = & -\log\left[-8\left|Z^{0}\right|^{2}y^{3}\right],\label{eq:K-STU}\\
K_{+} & = & -\log\left[-8\left|X^{0}\right|^{2}t^{3}\right],\label{eq:K+STU}\end{eqnarray}
are real, i.e. that $y,t<0.$ Comparing this
with \eqref{eq:ty-ex}, we arrive at a constraint on the signs of
the fluxes: $\mbox{sgn}\left(q_{x0}q_{0z}\right)=-1.$

We now solve for $y^{2}$ to find \begin{eqnarray}
y^{2} & = & -\frac{2}{3}\frac{q_{x0}q_{0z}}{q_{0}^{\hpa0}q_{x0}+q_{0z}q_{x}^{\hpa z}}\,.\end{eqnarray}
The denominator is superficially similar to the combination of fluxes
that appears in the tadpole constraint \eqref{eq:Tadpole-ex}, but
differs by a relative sign, ensuring that this solution is regular.
Indeed, we can impose the tadpole constraint to find two useful expressions
for $y^{2}:$\begin{equation}
y^{2}=-\frac{q_{0z}}{3q_{0}^{\hpa0}}=-\frac{q_{x0}}{3q_{x}^{\hpa z}}\,.\end{equation}
 Since $y$ is real, we can read off two more sign constraints: $\mbox{sgn}\left(q_{0z}q_{0}^{\hpa0}\right)=\mbox{sgn}\left(q_{x0}q_{x}^{\hpa z}\right)=-1.$
Combining these with the previous sign constraint, we conclude that
the allowed choices of signs for the fluxes are either $q_{0z},q_{x}^{\hpa z}>0$
and $q_{x0},q_{0}^{\hpa0}<0,$ or $q_{0z},q_{x}^{\hpa z}<0$ and $q_{x0},q_{0}^{\hpa0}>0.$
While it is possible that other sign choices lead to stable compactifications,
they cannot lead to solutions of the F-flatness conditions in the geometric regime.

After taking into account the Kähler cone constraints, the explicit
solutions for the moduli are \begin{eqnarray}
y & = & -\sqrt{-\frac{2}{3}\frac{q_{x0}q_{0z}}{q_{0}^{\hpa0}q_{x0}+q_{0z}q_{x}^{\hpa z}}}\,,\label{eq:y-sol-ex}\\
t & = & -\sqrt{-\frac{2}{3}\frac{\left(q_{0z}\right)^{3}}{q_{x0}}\frac{1}{q_{0}^{\hpa0}q_{x0}+q_{0z}q_{x}^{\hpa z}}}\,.\label{eq:t-sol-ex}\end{eqnarray}

\subsection{Mass Parameters}

In addition to determining the VEVs of the moduli $z^{i}$ and $x^{a},$ the
flux attractor equations also determine the Kähler-invariant quantities
$CX^{0}Z^{0}$ and $\widetilde{L}^{A\overline{I}}.$ These {}``mass
parameters'' appear in the mass matrices and Yukawa couplings of
the 4D effective theory. They are independent of the stabilized values
of the moduli, in the sense that we can find families of fluxes that
lead to the same $x$ and $z,$ but different values of the mass parameters.
We begin by computing $CX^{0}Z^{0},$ which determines the gravitino
mass, then move on to compute the $\widetilde{L}^{A\overline{I}}.$

If we recall \eqref{eq:WC-Real} and \eqref{eq:WC-Complex}, 
\begin{equation}
W= \frac{1}{2}e^{-K_{+}-K_{-}}\overline{C}=ie^{-K_{\pm}}\overline{C}_{\pm}\,,
\end{equation}
 we see that the gravitino mass can be written \begin{eqnarray}
e^{-K_{(S)}}m_{3/2}^{2} & = & e^{K_{+}+K_{-}}\left|W\right|^{2}\\
 & = & \frac{1}{4}e^{-K_{+}-K_{-}}\left|C\right|^{2}\\
 & = & e^{K_{\mp}-K_{\pm}}\left|C_{\pm}\right|^{2}.\end{eqnarray}
 Note that the dilaton is not stabilized by NS fluxes, so we have
used a rescaled gravitino mass $e^{-K_{(S)}}m_{3/2}^{2}$ which is
independent of the dilaton. Given the Kähler potentials \textbf{\eqref{eq:K-STU}
}-\eqref{eq:K+STU}, we find three equivalent expressions: \begin{eqnarray}
e^{-K_{(S)}}m_{3/2}^{2} & = & 64y^{3}t^{3}\left|CX^{0}Z^{0}\right|^{2}\\
 & = & \frac{y^{3}}{t^{3}}\left|\frac{C_{-}Z^{0}}{\overline{X}^{0}}\right|^{2}\\
 & = & \frac{t^{3}}{y^{3}}\left|\frac{C_{+}X^{0}}{\overline{Z}^{0}}\right|^{2}.\label{eq:grav-ex}\end{eqnarray}
 Given one of the three Kähler-invariants
$CX^{0}Z^{0},$ $C_{-}Z^{0}/\overline{X}^{0},$ and $C_{+}X^{0}/\overline{Z}^{0}$, 
we can use \eqref{eq:C+Def}, \eqref{eq:C-Def},
\eqref{eq:K-STU}, and \eqref{eq:K+STU} to determine the others.

In our example the easiest quantity to compute is $C_{+}X^{0}/\overline{Z}^{0},$
\eqref{eq:CX0-ex}. We first substitute \eqref{eq:y-sol-ex} into
the expressions for the complex fluxes \eqref{eq:eo-ex} and \eqref{eq:ex-ex},
then apply the tadpole constraint \eqref{eq:Tadpole-ex} to find \begin{eqnarray}
\frac{e_{x}^{+}}{Z^{0}} & = & -2q_{x0}\,,\\
\frac{e_{0}^{+}}{Z^{0}} & = & -\frac{2}{3}iyq_{0z}\,.\end{eqnarray}
 Substituting these into \eqref{eq:CX0-ex}, we find 
 \begin{equation}
\frac{C_{+}X^{0}}{\overline{Z}^{0}}=\frac{1}{6}i\frac{\left(q_{x0}\right)^{3}}{\left(q_{0z}\right)^{2}y^{2}}
 =\frac{i}{2}\left(-\frac{q_{x0}}{q_{0z}}\right)^{3}q_{0}^{\hpa0}
 \end{equation}
 and combine this with \eqref{eq:ty-ex} and \eqref{eq:grav-ex} to
find \begin{equation}
e^{-K_{\phi}}m_{3/2}^{2}=\frac{1}{4}\left(-\frac{q_{x0}}{q_{0z}}\right)^{3}\left(q_{0}^{\hpa0}\right)^{2}\,.\end{equation}
The righthand side is also equal to $\mathcal{G}_\pm/2$ at the attractor point, as noted in \eqref{eq:G-gravitino}.

While the gravitino mass can be computed using only the projected
attractor equations, we return to the full set of attractor equations
\eqref{eq:RealAtt-SU3-1},~\eqref{eq:RealAtt-SU3-2} to compute the
$\widetilde{L}^{A\overline{I}}.$ While the projected attractor equations
were non-linear in the moduli, the remaining attractor equations are
linear in the $\widetilde{L}^{A\overline{I}},$ with solutions \begin{eqnarray}
\mbox{Im}\left\{ \widetilde{L}^{B\overline{I}}\right\}  & = & \frac{1}{2}\mbox{Im}\left\{ G_{AB}\right\} ^{-1}q_{A}^{\hpa I}\,,\label{ImL}\\
\mbox{Re}\left\{ \widetilde{L}^{B\overline{J}}\right\}  & = & \frac{1}{2}\mbox{Im}\left\{ G_{AB}\right\} ^{-1}\left(q_{AJ}+q_{A}^{\hpa K}\mbox{Re}\left(F_{KJ}\right)\right)\mbox{Im}\left(F_{IJ}\right)^{-1}.\label{ReL}\end{eqnarray}
 We can compute the moduli-dependent matrices, \begin{eqnarray}
\mbox{Im}\left\{ G_{AB}\right\} ^{-1} & = & \left(\begin{array}{cc}
-1/2t^{3} & 0\\
0 & 1/6t\end{array}\right),\\
\mbox{Im}\left\{ F_{IJ}\right\} ^{-1} & = & \left(\begin{array}{cc}
-1/2y^{3} & 0\\
0 & 1/6y\end{array}\right),\\
\mbox{Re}\left\{ F_{IJ}\right\}  & = & 3t^{2}\left(\begin{array}{cc}
0 & 1\\
1 & 0\end{array}\right),
\end{eqnarray}
 and determine \eqref{ImL} and \eqref{ReL} as 
 \begin{equation}
\mbox{Im}\left\{ \widetilde{L}^{B\overline{I}}\right\}   =  \frac{1}{4}q_{x}^{\hpa z}\sqrt{-3\frac{q_{x0}q_{x}^{\hpa z}}{\left(q_{0z}\right)^{2}}}\left(\begin{array}{cc}
-3q_{0}^{\hpa0}q_{x0}/\left(q_{0z}\right)^{2} & 0\\
0 & 1/3\end{array}\right),
\end{equation}
\begin{equation}
\mbox{Re}\left\{ \widetilde{L}^{B\overline{J}}\right\}  = \frac{1}{24}\left[1-\left(\frac{q_{0z}}{q_{x0}}\right)^{2}\right]\left(\frac{3q_{x}^{\hpa z}}{q_{0z}}\right)^{2}\left(\begin{array}{cc}
0 & q_{x0}\\
q_{0z} & 0\end{array}\right).
\end{equation}
We made use of the tadpole constraint \eqref{eq:Tadpole-ex}
and the solutions for the moduli \eqref{eq:y-sol-ex} and \eqref{eq:t-sol-ex}
to simplify these expressions. Assembling the results, we have \begin{eqnarray}
\widetilde{L}^{0\overline{0}} & = & -\frac{3}{4}i\frac{q_{x}^{\hpa z}q_{0}^{\hpa0}q_{x0}}{\left(q_{0z}\right)^{2}}\sqrt{-3\frac{q_{x0}q_{x}^{\hpa z}}{\left(q_{0z}\right)^{2}}}\,,\\
\widetilde{L}^{0\overline{z}} & = & \frac{1}{24}\left(\frac{3q_{x}^{\hpa z}}{q_{0z}}\right)^{2}q_{x0}\left[1-\left(\frac{q_{0z}}{q_{x0}}\right)^{2}\right]\,,\\
\widetilde{L}^{x\overline{0}} & = & \frac{1}{24}\left(\frac{3q_{x}^{\hpa z}}{q_{0z}}\right)^{2}q_{0z}\left[1-\left(\frac{q_{0z}}{q_{x0}}\right)^{2}\right]\,,\\
\widetilde{L}^{x\overline{z}} & = & \frac{i}{12}q_{x}^{\hpa z}\sqrt{-3\frac{q_{x0}q_{x}^{\hpa z}}{\left(q_{0z}\right)^{2}}}\,.\end{eqnarray}
Along with the VEVs \eqref{eq:y-sol-ex} and \eqref{eq:t-sol-ex} these constitute explicit solutions to the attractor equations for the simple prepotentials \eqref{eq:F-ex} and \eqref{eq:G-ex}.

\acknowledgments
LA thanks the Michigan Center for Theoretical Physics for hospitality, where this work was initiated.  The work of FL and RO was supported by
the DOE under grant DE-FG02-95ER40899. The research of LA is supported by DOE grant FG02-84-ER40153.

\end{document}